\documentclass[%
 reprint,
 amsmath,amssymb,
 aps,
]{revtex4-2}

\usepackage{color}

\usepackage{graphicx}
\usepackage{dcolumn}
\usepackage{bm}
\usepackage{hyperref}
\usepackage[mathlines]{lineno}
\usepackage{float}
\usepackage{subfig}

\begin{document}

\markboth{Matheus C. Teodoro, Lucas G. Collodel 
and Jutta Kunz}
{Tidal effects in the motion of gas clouds around boson stars}

\title{Tidal effects in the motion of gas clouds around boson stars}

\author{Matheus C. Teodoro}
\address{Institute of Physics, University of Oldenburg\\
26111 Oldenburg, Germany\\
matheus.do.carmo.teodoro@uol.de}

\author{Lucas G. Collodel}
\address{Theoretical Astrophysics, University of T\"ubingen\\
72076 T\"ubingen, Germany\\
lucas.gardai-collodel@uni-tuebingen.de}

\author{Jutta Kunz}
\address{Institute of Physics, University of Oldenburg\\ 26111 Oldenburg, Germany\\
jutta.kunz@uni-oldenburg.de}


\begin{abstract}

We study the motion of gas clouds in the vicinity of boson
stars, performing simulations with the Black Hole Accretion
Code BHAC. We compare the motion of the gas clouds with
particle motion along geodesics and analyze the tidal
effects on the gas clouds, which leads to disruption of the
clouds.
First we consider small and dense clouds associated with
three different types of bound orbits close to the boson
star and analyze the mechanisms of debris formation for
these. We infer from the simulations that the lifetimes 
of these nearby clouds are longer for initially circularly 
orbiting clouds than for clouds on initially eccentric
orbits. 
Next we compare the evolution of more extended 
and less dense clouds on initially circular orbits 
around a boson star and a Schwarzschild black hole
and compare the motion in these two spacetimes.
In particular, we observe the formation of a ring-like
structure around the boson star 
endowed with a spiralling shock structure 
and a constant thermal bremsstrahlung total luminosity. 
This final configuration contrasts strongly with the 
black hole scenario where the gas is totally captured
behind the event horizon.   

\end{abstract}

\maketitle  

\section{Introduction}

In recent years black holes have featured most
prominently in astrophysical observations.
On the one hand, the LIGO/VIRGO collaboration
has detected gravitational waves produced in the
powerful merger events of stellar black holes, 
giving even rise to an intermediate size black hole
(see e.g. \cite{Abbott:2016blz,Abbott:2016nmj,Abbott:2017vtc}).
On the other hand, the EHT collaboration
has presented observations of the shadow
and the accretion disk for the supermassive compact object 
at the center of the galaxy M87
\cite{Akiyama:2019cqa,Akiyama:2019brx,Akiyama:2019sww,Akiyama:2019bqs,Akiyama:2019fyp,Akiyama:2019eap}.
Notwithstanding, over the course of two decades the MPE and UCLA teams have collected data, observing stars' motion in the vicinity of Sagittarius A*, which strongly suggests that at the center of our Milky Way galaxy dwells a supermassive black hole. 
\cite{Ghez:1998ph,Ghez:2003qj,Genzel:2003as,Ghez:2008ms,Gillessen:2008qv,Genzel:2010zy,Eckart:2017bhq}.

Still, the case for possible contenders
of black holes is not yet closed, 
leaving the need to scrutinize these further.
Exotic compact objects warp spacetime in such a manner 
that it is regular everywhere 
and no horizon is formed. 
However, since they may possess a high compactness,
ergospheres and photon spheres may arise, 
letting these objects act as black hole mimickers. 
For instance, the gravitational
wave spectrum of the ringdown phase of exotic
compact objects can be initially almost identical
to the one of a black hole 
\cite{Cardoso:2016rao,Cardoso:2016oxy}.

First studied in the late 60's 
\cite{Feinblum:1968nwc,Kaup:1968zz,Ruffini:1969qy}
boson stars (BSs) are formed by a complex scalar field
bound by gravity, 
and hence BSs seem a rather attractive candidate for 
exotic compact objects.
In fact, their sizes can range from the atomic scale 
to the scale of supermassive
black holes (BHs). 
Thus such a supermassive BS might well dwell 
at the center of our very own galaxy 
\cite{Torres:2000dw,Berti:2006qt,Vincent:2015xta}.
Albeit the simplicity of this type of matter fields, 
the resulting physics is extremely interesting, 
mainly because of the stars' features 
that differ enormously 
from those of other final state systems 
such as BHs or neutron stars. 
The generated spacetime is asymptotically flat, 
but the scalar field becomes only trivial 
at spatial infinity,
and the star possesses therefore no clear boundary
that could be identified with its surface 
where the pressure vanishes. 
Yet the investigation of BSs as realistic contenders 
of astrophysical relevance lies in the fact 
that the complex scalar field only 
interacts gravitationally with ordinary matter, 
which can freely orbit in the stars' interior 
all the way to the core with no resistance 
from the bosonic field.

The theory underlying the existence of BSs 
is endowed with a global $U(1)$ symmetry,
and thus possesses a conserved Noether current 
and associated charge $N$, 
understood as the particle number. 
It is also possible to promote this symmetry 
to a local one by gauging the scalar field, 
which then sources an electromagnetic-like field
\cite{Jetzer:1989av}.
Rotating BS solutions were first obtained in the 90's,
after the perturbative approach proved to be of no avail
\cite{Kobayashi:1994qi}.
 The underlying reason is that the angular momentum 
of rotating BSs
is quantized in units of the particle number, $J=mN$ 
(with $m\in\mathbb{Z}$) 
\cite{Mielke:1997ag,Yoshida:1997qf}.
Solving the full nonlinear set of partial differential
equations numerically, numerous sets of rotating BSs 
were obtained and analyzed, including their stability, excitations, existence also in higher dimensions and
generalizaton to multistate BSs, reflecting the richness 
of their nature in a plethora of different configurations 
\cite{Friedberg:1986tq,Ryan:1996nk,Schunck:1996he,Schunck:1999pm,Kleihaus:2005me,Kleihaus:2007vk,Bernal:2009zy,Hartmann:2010pm,Kleihaus:2011sx,Collodel:2019ohy,Li:2019mlk}.

Geodesic motion of test particles in BS spacetimes 
has been studied in 
\cite{Eilers:2013lla,Macedo:2013jja,Grandclement:2014msa,Meliani:2015zta,Grandclement:2016eng,Grould:2017rzz,Collodel:2017end}, 
where unusual types of orbits have been found, 
not present in Schwarzschild or 
Kerr BH spacetimes,
such as the \emph{semi orbit}, 
the \emph{pointy petal orbit} 
and the \emph{static ring}.
The latter represents a set of orbits,
where a particle at rest with respect to an asymptotic static observer remains at rest in a static orbit
\cite{Collodel:2017end}.
The investigation of the circular orbits
of particles in BS spacetimes has also revealed 
distinct features as compared to BH spacetimes.
Whereas BH spacetimes feature ISCOs,
i.e., innermost stable circular orbits, 
where a change of stability occurs,
boson stars feature ICOs,
i.e., innermost circular orbits, 
beyond which no circular motion is possible
\cite{Grandclement:2014msa,Meliani:2015zta,Teodoro:2020kok}.

Circular orbits are of particular importance in the
study of accretion disks around compact objects.
Regarding thick tori around BSs, 
Meliani et al.~\cite{Meliani:2015zta} have reported 
important differences between those 
in the BS and BH context, 
by exploring analytical solutions 
of stable circular fluid configurations 
and their evolution through simulations. 
Olivares et al.~\cite{Olivares:2018abq} 
have explored magnetized tori 
together with general-relativistic 
radiative-transfer calculations,
pointing out also potential differences 
in the appearance of BSs and BHs. 
Also, the iron K$\alpha$ line of thin accretion disks 
around mini BSs has been analyzed \cite{Cao:2016zbh} 
and found to be in most cases incompatible 
with current x-ray data from BH binaries. 
However, compatibility with the data
might change dramatically for compact BSs,
composed of self-interacting fields.

The motion of gas clouds around rotating BSs has been
studied in \cite{Meliani:2017ktw}, 
where simulations have been performed 
for near-by and far away clouds.
Under the influence of the gravitational field
from the BSs, these clouds have been
tidally distorted and disrupted.
For the two regimes of near-by and far away clouds
major differences were found.
While for clouds falling into the BSs from 
a great distance the formation of a 
torus-like structure inside the BSs was reported, 
the near-by clouds were capable to retain 
largely their shape.

In the present paper we shall also address 
the motion and distortion of gas clouds,
but in the context of non-rotating compact BSs. 
Unlike the clouds considered in \cite{Meliani:2017ktw}, 
the clouds reported here also do not necessarily 
start at rest with respect to a zero angular momentum
observer at infinity. 
In fact, without the Lense-Thirring effect, 
in order to explore angular motion of the gas, 
initial angular momentum of the clouds is required. 
Our simulations are then performed in two regimes, 
consisting of dense small clouds with orbits 
close to the BSs 
and more extended and less dense clouds 
further away from the BSs. 
The former allows us to distinguish between the 
gas motion and the particle geodesic motion, 
thereby finding the debris formation mechanism, 
while the latter is meant to compare the evolution 
around a BS with that of a BH, as well
as to investigate the possibility for disk formation.
In both regimes we have considered the total mass 
of the clouds to be much smaller than the mass 
of the central compact object. 
This assumption obviates the necessity of defining 
a tidal radius since the self-gravity of the clouds 
will not be considered. 

Following an analogous approach to 
Meliani et al.~\cite{Meliani:2017ktw}, 
we restrict our analysis to the 2D case. 
Accordingly, the thickness of the cloud is neglected, 
as well as any component of the dynamical quantities
orthogonal to the cloud's plane. 
Such a restriction is compatible with the nature 
of the spherically symmetric spacetimes analysed,
and valid as a preliminary approach to the 
dynamics of the clouds. 
Indeed, important features such as the vertical mode 
of the magneto rotational instability (MRI) turbulence 
are irrelevant since we are not considering 
the presence of magnetic fields. 
Also, although vertical velocity components would
contribute to the pressure oscillations observed, 
such contributions are not expected to divert our results, 
when considering the gas to be symmetric 
with respect to the equatorial plane chosen. 

This paper is structured in the following format. 
In sections 2 and 3 we shall describe, respectively, 
the BS model employed 
and the numerical method together 
with the simulation setup. 
Sections 4 and 5 describe and discuss the simulations,
while section 6 presents our conclusions.

\section{Boson star model}

Boson stars are obtained by minimally coupling a complex scalar field 
to gravity. 
The action of the system is given by

\begin{eqnarray}
S=\int \Big[\frac{\cal{R}}{16 \pi G}
-\frac{1}{2}g^{\mu\nu}\left(\partial_{\mu}\Phi\partial_{\nu}\Phi^{*}+\partial_{\nu}\Phi\partial_{\mu}\Phi^{*}\right)\nonumber\\
-U\left(\lvert\Phi\rvert\right)\Big] \sqrt{-g}d^4x,
\label{bsaction}
\end{eqnarray}

where ${\cal R}$ is the curvature scalar, $G$ is Newton's constant,
$g^{\mu\nu}$ is the inverse metric, 
$\Phi$ is the complex scalar field, 
$U$ is the self-interaction potential, and $g$ is the metric determinant. 
The model has a U(1) invariance with conserved current
$j^\mu = -i( \Phi^{*} \partial^\mu \Phi - \Phi \partial^\mu  \Phi^{*})$. Boson stars could form thanks to a phenomenon known as \emph{gravitational cooling},
which is dissipationless and similar to the violent relaxation of collisionless stellar systems,
but yet more efficient, i.e. even for high ratios between kinetic and potential energy a compact object
is formed by ejecting scalar material carrying away the exceeding energy, and this final state is quite
insensitive to the initial conditions \cite{Seidel:1993zk,Sanchis-Gual:2019ljs}.

We vary the action with respect to the metric and the scalar field 
to obtain the Einstein field equations and a Klein-Gordon equation, 
respectively. 
We then employ the usual spherically symmetric ansatz,
where the scalar field has
a harmonic time-dependence $\Phi=\phi(r)\exp{(i \omega_s t)}$, and
the line element reads
\begin{equation}
\label{ds2}
ds^2=-e^{f(r)}dt^2+e^{l(r)}dr^2+r^2d\Omega^2,
\end{equation}
to obtain a system of three coupled ODEs, namely
\begin{equation}
\label{eql}
l'=\phi'^2r^2+ e^{l-f}\omega_s^2\phi^2r+\frac{1}{r}\left(e^l-1\right)- e^l Ur,
\end{equation}
\begin{equation}
\label{eqf}
f'=\phi'^2r^2+ e^{l-f}\omega_s^2\phi^2r-\frac{1}{r}\left(e^l-1\right)+ e^l Ur,
\end{equation}
\begin{equation}
\label{eqphi}
\phi''=\phi'\left(\frac{l'}{2}-\frac{f'}{2}-\frac{2}{r}\right)+\frac{1}{2}e^l\partial_\phi U-e^{l-f}\omega^2\phi,
\end{equation}
where the primes correspond to the derivatives with respect to the radial coordinate and $\partial_\phi\equiv\partial/\partial_\phi$, and adopt the appropriate boundary conditions which ensure regularity and asymptotic flatness. 
The numerical solutions are obtained with the 
aid of the program package Colsys \cite{Ascher:1979iha}. 
This solver uses a collocation method for boundary-value ODEs together with 
a damped  Newton method of quasi-linearization. 
The linearized problem is solved at each iteration step, by using a spline collocation at Gaussian points. 
The package is able to adapt the mesh using a selection procedure, 
in which the equations are solved on a sequence of refined
meshes until some stopping criterion is reached,
specifying the error of the numerical solutions.
The relative precision for the functions obtained is typically below $10^{-10}$.

The resulting spacetime then serves as the stage 
for the simulations
of the motion and evolution
of the gas clouds.
The potential $U$ defines the self-interaction
of the scalar field and includes the mass term for the scalar field.
For the \emph{mini BSs} the potential
has only a mass term, $U=m_b^2\lvert\Phi\rvert^2$,
where $m_b$ denotes the mass of the bosons.
Mini BSs do not possess high compactness.
Neither can they reach high total masses,
unless the boson mass $m_b$ is extremely small.
For the \emph{solitonic BSs}, on the other hand, for which a sextic
self-interaction potential is employed,
\begin{equation}
\label{bspot}
U\left(\lvert\Phi\rvert\right)
=\lvert\Phi\rvert^2\left(m_b^2+a\lvert\Phi\rvert^2+b\lvert\Phi\rvert^4\right),
\end{equation}
very high compactness close to the BH limit can be achieved.
(For further details on various BS properties we refer the reader
to the most recent review on the topic \cite{Liebling:2012fv}.)

\begin{figure}[t!]
  \centering
  \hspace{1cm}{\includegraphics[width=1\columnwidth]{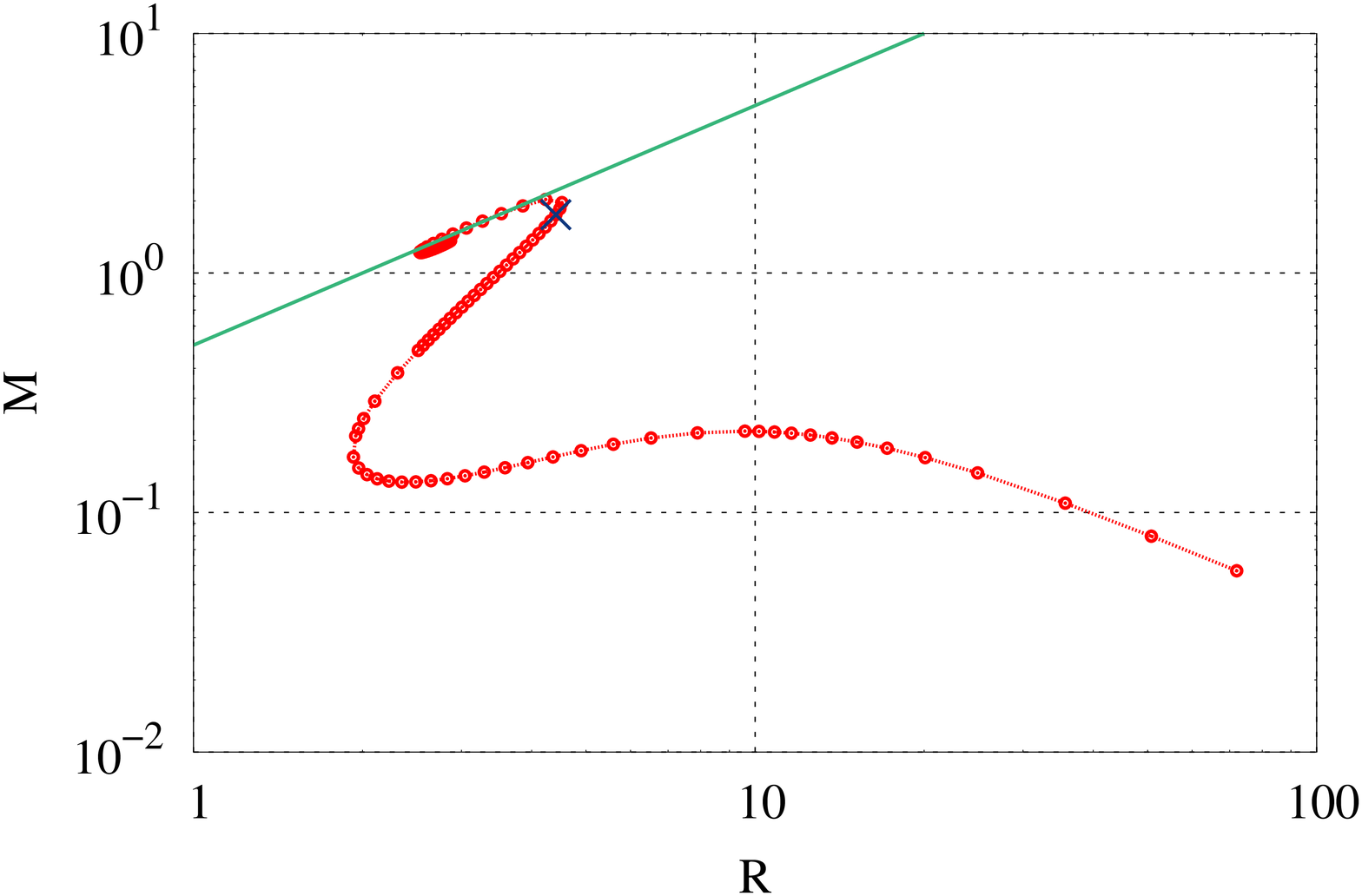}}
  {\includegraphics[width=1.1\columnwidth]{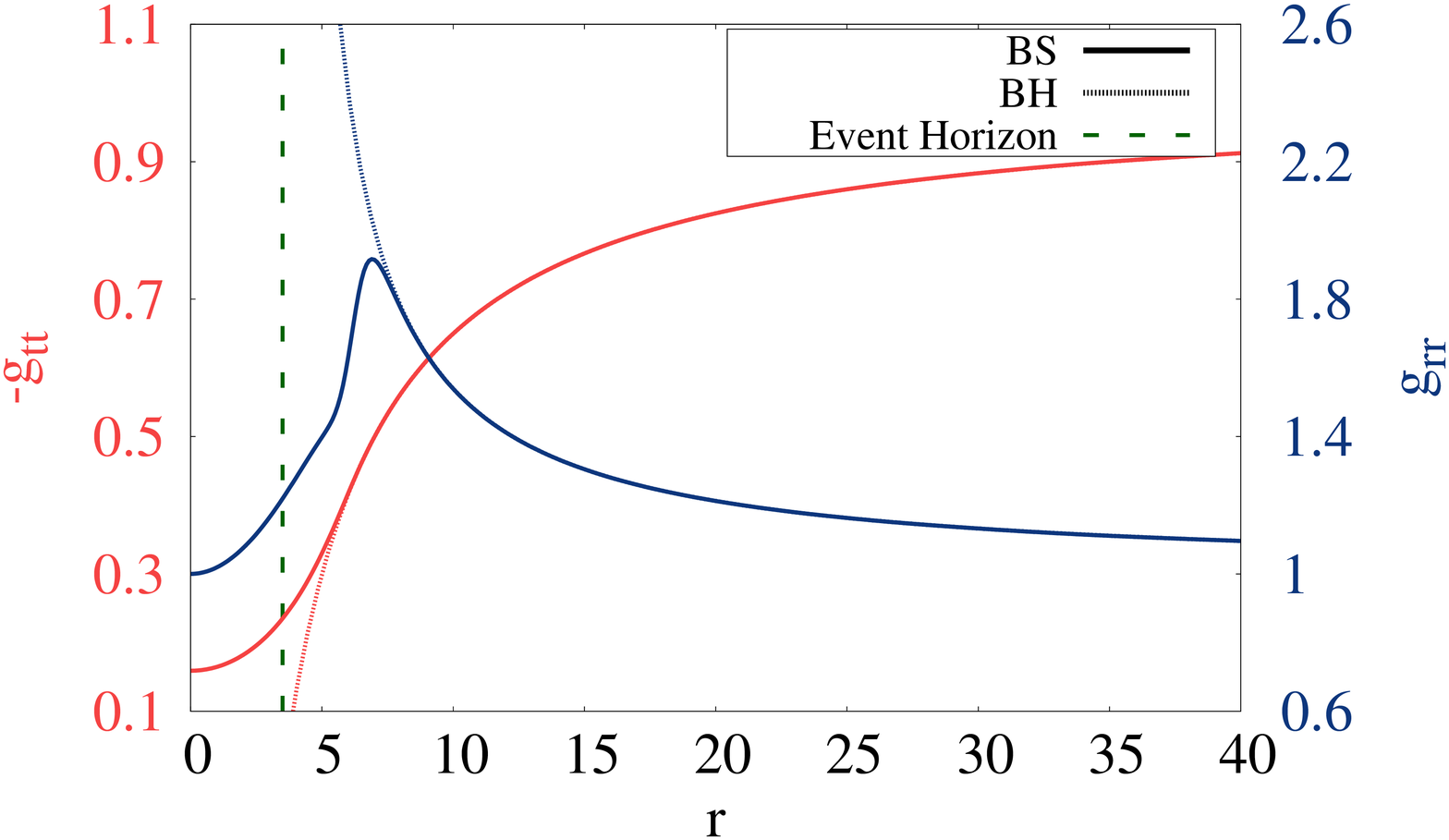}}

\caption{(a - Upper panel) Mass $M$ vs radius $R$ relation for a set of compact
BSs ($R=R_{\rm BS}$, red) and Schwarzschild BHs ($R=R_{\rm H}$, green).
The BS employed in the simulations is also indicated (blue cross) at $R_{BS}=4.42$.
(b - Lower panel) Metric function $-g_{tt}$ (left scale) of the selected BS (red)
and the Schwarzschild BH (light red) vs the radial coordinate $r$,
together with the metric function $g_{rr}$ (right scale)
of the selected BS (blue) and the Schwarzschild BH (light blue).
The event horizon radius $R_{\rm H}$ is also indicated (black).
}
     \label{mxr_bs}
\end{figure}

In Fig.~\ref{mxr_bs}(a) we present the mass-radius relation
for a set of solutions of solitonic BSs.
(Here the parameters of the potential $U$ are chosen as 
$m_b=1$, $a=-2$, $b=1$; 
the gravitational coupling is $\kappa=8 \pi G {m_b^2}/{b} =0.1$.)
Since BSs possess no sharp surface, a common approach employed here
is to define the BS radius 
via an integral over the particle number density $j^t=-2e^{-f}\omega_s\phi^2$,
\begin{equation}
R_{\rm BS}= \frac{\int j^t \sqrt{-g}\, r dr }
     {\int j^t \sqrt{-g}\, dr },
\label{radius}
\end{equation}
weighted with the Schwarzschild-like radial coordinate $r$,
and normalized with respect to the total particle number
\cite{Schunck:2003kk,Kleihaus:2011sx}.
This definition must nevertheless be taken with a grain of salt,
since although it provides us with a measure of compactness,
it differs from the usual one.
This is also seen in the diagram,
where the green line represents the Schwarzschild BH 
mass-radius relation.
Whereas some BS solutions lie just above the green line,
by no means this is to be interpreted
as the solutions being confined to a radius
smaller than their Schwarzschild radius.

We also indicate in the figure the BS chosen in our simulations
(blue cross, obtained with boson frequency $\omega_s=0.27$). This particular
solution has a mass of $M_{BS}=1.75$ and a radius of $R_{BS}=4.42=2.53M_{BS}$.
Thus it has a high compactness,
$C_{BS}=M_{BS}/R_{BS}=0.4$, which is
close to the compactness of a Schwarzschild BH, $C_{BH}=0.5$.

The metric functions $g_{tt}$ (left scale, red)
and $g_{rr}$ (right scale, blue) of this star
are illustrated in Fig.~\ref{mxr_bs}(b)
together with the metric functions of a Schwarzschild BH
with the same mass. The event horizon radius of the BH
is also indicated.
Note that
this BS solution belongs to the stable BS branch,
residing below the BS solution of maximum mass, where stability is lost.

\section{Numerical methods}

We have performed the simulations using BHAC \cite{Porth:2016rfi,Olivares:2019dsc}, 
evolving the ideal general relativistic fluid equations 
in the inviscid fluid limit. 
In the following we briefly recall the set of equations
to be solved and discuss the numerical setup. Geometric coordinates are used, $G=c=1$,
and the normalization mass is set to one ($M=1$)
for the BS or BH representing the compact object,
where $M_{\rm BH}=M_{\rm BS}=1.75M$. 

\subsection{Evolution equations}

The evolution equations, 
providing mass and stress energy momentum conservation, 
read in covariant notation,
\begin{equation}
\begin{split}
\nabla_\mu (\rho u^{\mu})=0,\\
\nabla_\mu T^{\mu\nu}=0,
\end{split}
\label{MainEq}
\end{equation}
where $\rho$ is the rest-mass density of the fluid and $u^\mu$ the four-velocity.
The stress energy tensor of the fluid reads
\begin{equation}   
T^{\mu\nu}=\rho h u^\mu u^\nu +p g^{\mu\nu}.
\end{equation}
Here $p$ is the pressure of the fluid,
and the enthalpy $h$ will be defined by its equation of state (EOS).
Considering an ideal gas, we have
\begin{equation}       
h(p,\rho)=1+\frac{\hat{\gamma}}{\hat{\gamma}-1}\frac{p}{\rho},  
\end{equation}
where $\hat{\gamma}$ is the adiabatic index. 
In the model we are considering 
a non-degenerate non-relativistic fluid, 
thus $\hat{\gamma}=\frac{5}{3}$. 

Similarly to the Valencia formalism \cite{RezzollaBook} 
we proceed with the $3+1$ decomposition of the spacetime,
defining the four-velocity of the {\it Eulerian observer} 
as the unit vector normal to the space-like foliation $\Sigma_t$,
defined as the iso-surfaces with respect to the scalar time function $t$
\begin{equation}
n_\mu=-\alpha\nabla_\mu t,
\end{equation}
with $\alpha$ being the {\it lapse} function. 
Considering a stationary spherically symmetric metric, 
the line element in such a decomposition reads
\begin{equation}
ds^2=-\alpha^2dt^2+\gamma_{ij}{dx^idx^j},
\end{equation}
where $\gamma_{ij}$ is the full spatial metric on the iso-surfaces. 
Thus the three-velocity of the fluid can be found with the projection
\begin{equation}
v^i=\frac{\gamma^i_{\mu}u^\mu}{\Gamma}=\frac{u^i}{\Gamma}, 
\end{equation}{}
where $\Gamma=u^\mu n_\mu $ is the Lorentz factor, 
which reduces to $\sqrt{1-v^2}$.

Within this framework we aim to evolve the quantities $\rho$, $p$ and $v^i$. 
On the other hand, in order to rewrite eqs.~(\ref{MainEq})
in the conservative form 
\begin{equation}
    \partial_t\mathbf{U}+\partial_i\mathbf{F}^i=\mathbf{S},
\end{equation}
one must define the conservative variables
\begin{equation}
    \mathbf{U}:=
    \sqrt{g_s}
    \begin{pmatrix}
     D\\
     S_j\\
     \tau
    \end{pmatrix}
    =\sqrt{g_s}
    \begin{pmatrix}
    -\rho \Gamma\\
    \rho h \Gamma^2 v_j\\
    \rho h \Gamma^2 -p - D
    \end{pmatrix},
\end{equation}
where $g_s$ is the determinant of the spatial metric, $g_s=\det[\gamma_{ij}]$.
Then the conserved variables in the Eulerian frame are 
the density $D=\rho\Gamma$, the covariant three-momentum 
$S_i=\rho h\Gamma^2v_i$ and the energy density 
$\tau=\rho h \Gamma^2-p-D$. 
For these quantities, the fluxes read
\begin{equation}
    \mathbf{F}^i=
    \sqrt{g_s}
    \begin{pmatrix}
    D\alpha v^i\\
    \alpha W^i_j  \\
    \alpha(S^i-v^i D)
    \end{pmatrix},
\end{equation}
and the source terms are
\begin{equation}
    \mathbf{S}=
    \sqrt{g_s}
    \begin{pmatrix}
      0\\
      \frac{1}{2}\alpha W^{ik}\partial_j\gamma_{ik}-U\partial_j\alpha\\
      -S^j\partial_j\alpha
    \end{pmatrix},
\end{equation}
with the spatial stress energy tensor 
$W_{ij}=ph\Gamma^2 v_iv_j -p\gamma_{ij}$. 
Through BHAC the spatial splitting of these equations is done 
using the total variation diminishing Lax-Friedrichs method 
combined with a piece-wise parabolic limiter 
and the time integration through an order predictor-corrector type 
``twostep'' scheme \cite{Porth:2016rfi}.

In order to perform 2D simulations we restrict the above equations to the equatorial plane, 
$\{x^1,x^2,x^3\}\rightarrow\{r,\phi,\theta=\pi/2\}$, 
employing Schwarzschild-like spherical coordinates. 
We then take into account that the three-velocity, flux and source term in the $x^3-$direction must vanish. 
Therefore we evolve the above conservation laws by taking the Roman indices range to be $\{1,2\}$. 
For the simulations in which the gas cloud crosses the core of the star, 
we apply a Cartesian grid and transform the tensor components accordingly.

From the hydrodynamic quantities one can also estimate 
the temperature $T$ of the ideal gas through
\begin{equation}
T=\frac{p}{\rho}.
\end{equation}
The sound speed $c_s$ and the relativistic Mach number $\mathcal{M}$ read
\begin{equation}
\begin{split}
c_s^2= \frac{\hat{\gamma}(\hat{\gamma}-1)p}{\rho(\hat{\gamma}-1)+\hat{\gamma}p},\\
\mathcal{M}=\frac{(v^iv_i)^{\frac{1}{2}}\Gamma}{c_s\Gamma_{c_s}}, 
\end{split}
\end{equation}
where $\Gamma_{c_s}=1/\sqrt{1+c_s^2}$ is
the Lorentz factor of the sound speed.
For the thermal Bremsstrahlung emissivity $\varepsilon$ 
a simplified assumption is employed
\cite{LightmanBook} 
\begin{equation}
\varepsilon=T^{1/2}\rho ,
\end{equation}
considering that the gas is hot and ionized 
and taking into account that we do not perform radiation transfer 
in our simulations. 

Global variables also are computed, in particular, 
the total luminosity $L$ and the mass flux $\dot{M}$
\begin{equation}
\begin{split}
L=\int\epsilon\Gamma\sqrt{g_s} drd\phi,\\
\dot{M}=-\int_{R_{\rm BS}}\rho \Gamma v^{r}\sqrt{g_s}d\phi.
\end{split}
\end{equation}
The surface integral calculated in order to obtain $\dot{M}$, is performed on circles of radius 
$R_{\rm BS}$, eq.~(\ref{radius}).

The maximum density $\rho_{\rm max}$ and 
the maximum pressure $p_{\rm max}$ 
are also calculated at each time-step. 
They are important variables since they indicate 
compression and expansion of the gas 
as well as possible stationary final configurations of it, when constant.
Furthermore, in order to capture shock waves reliably
we also apply the shock wave detector of the type described in \cite{Zanotti:2010xs}.

The detector operates by comparing the hydrodynamics quantities in two adjacent cells to predict if a shock front will occur. Here we will not provide a full description of the detector but only illustrate how it works. Restricting to one direction of propagation, the criterion for a shock wave to occur between two adjacent cells $1$ and $2$ is given by  
\begin{equation}
   V_{12}>\tilde{V}_{12}:=\tanh{\left(
   \int^{p_1}_{p_2}\frac{\sqrt{h^2+\mathcal{A}^2_1(1-c_s^2)}}{(h^2+\mathcal{A}_1^2)\rho c_s}
   dp \right)} ,
\end{equation}
where $\mathcal{A}_1:=h_1\Gamma_1 v_1$ (for a 2D case only), and $V_{12}:=(v_1-v_2)/(1-v_1v_2)$ . The threshold can be found by obtaining the limiting relative velocity for a shock-rarefaction pattern to occur. In order to quantify this condition in an output, we define the shock detector output to be
\begin{equation}
    S_d=\max\{0,V_{12}^x-\tilde{V}_{12}^x,V_{12}^y-\tilde{V}_{12}^y \},
\end{equation}
where now the superscripts $x$ and $y$ represent the two directions of propagation for the Cartesian grid. More details and the derivation of this type of detector can be found in \cite{RezzollaBook,Zanotti:2010xs}.


\subsection{Simulation setup}

The simulations have been performed on a 2D grid 
in the equatorial plane of the spacetimes explored. 
As initial condition the clouds have a Gaussian density distribution,
possessing a standard deviation $R$, centered 
at the radial coordinate $r_0$ with a maximum value 
of the density, $\rho_0$. 
The clouds start in thermal equilibrium with the medium,
meaning that the pressure is initially constant 
over the entire grid and set equal to $p=10^{-7}$. 
A constant non-vanishing angular momentum $u_\phi=\mathbf{L}$ 
is given to the cloud in some of the simulations. 
Table \ref{tab:table1} contains the details 
of each simulation presented, labeled S1 -- S5.
The maximum density of the denser clouds simulated here, $\rho_0$,
can also be taken as a normalization constant 
for the pressures and densities. Thus pressures and densities presented in this paper should be taken as fractions of $\rho_0$. It is important to mention that this constant can only be chosen consistent with our assumption that the total mass of the clouds is much smaller than the total mass of the BSs/BHs.
Together with the geometrized coordinates used,
and the central compact object's mass employed,
this normalization constant makes all the results 
presented here dimensionless. 

All the simulations presented have been performed such as to
allow four refinement levels. Except for the simulation S5
the grids have been chosen to be cartesian 
in order to avoid numerical problems at the origin. 
S5, in contrast, has been performed on a polar grid 
in order to avoid numerical problems near the event horizon. 

\begin{table*}[t!]
 {\footnotesize
    \begin{tabular}{|c|c|c|c|c|c|c|c|c|c|c|c} 
      \hline
     \textbf{Label} &  \textbf{Metric} & \textbf{Grid type} & \textbf{Grid size} & \textbf{$1^{th}$ level resolution} & $r_0$ & $R$ & $\mathbf{L}$ & $\rho_{0}$ \\
      \hline
      S1 & BS & cartesian & $[-7,7]\times[-7,7]$ & $[N_x,N_y]=[128,128]$ & 4 & 0.03 & 0 & $10^{ 0}$ \\
      \hline
      S2 & BS & cartesian & $[-7,7]\times[-7,7]$ & $[N_x,N_y]=[128,128]$ & 4 & 0.03 & 1.789 & $10^{ 0}$ \\
      \hline
      S3 & BS & cartesian & $[-7,7]\times[-7,7]$ & $[N_x,N_y]=[128,128]$ & 4 & 0.03 & 3.583 & $10^{ 0}$ \\
      \hline
      S4 & BS & cartesian & $[-30,30]\times[-30,30]$ & $[N_x,N_y]=[128,128]$ & 10  & 0.3 & 6.075 & $10^{-2}$\\
      \hline
      S5 & BH & polar & $r_{\rm min}=4, r_{\rm max}=30$ & $[N_r,N_{\phi}]=[64,256]$  & 10 & 0.3 & 6.075 & $10^{-2}$ \\
      \hline
    \end{tabular}
    }
    \caption{Simulation setup: The Cartesian grids are all square grids 
while the polar grid ranges from $r_{min}$ to $r_{max}$ 
in the radial coordinate and from $0$ to $2\pi$ in the angular coordinate.
$N_x$, $N_y$, $N_r$, and $N_{\phi}$ are the number of cells 
in the first level of refinement in the $x$, $y$, $r$ and $\phi$ coordinates.
$r_0$ represents the initial location of the center of the cloud,
$R$ its standard deviation, $u_\phi=\mathbf{L}$ its angular momentum,
and $\rho_0$ its maximum density.
}
        \label{tab:table1}
\end{table*}

As in any finite volume hydrodynamics simulation 
an atmospheric treatment is required.
We have chosen the atmosphere to be isotropic, static ( $v^r=v^\phi=0$ )
and rarefied  with values for density and pressure $\rho_{atm}=10^{-6}$ and $p_{atm}=10^{-7}$. 
The need of imposing a static atmosphere is 
to avoid early atmospheric accretion 
and to ensure initial thermal equilibrium 
between the cloud and the medium. 
On the other hand, in order to relax such a strong imposition,
we have also provided the region around the cloud with a tracer. 
The tracer, a scalar quantity advected with the fluid,
has initial value $1$ in the region surrounding the cloud 
and vanishes on the rest of the grid. 
The atmospheric values are then set only in the regions,
where the tracer is smaller than $10^{-7}$. 
The global variables are then calculated 
by taking only those cells into account 
for which the atmospheric condition does not apply.

We note that the Courant-Friedrichs-Lewy (CFL) constraint was also applied for the 
evolution of the equations, with CFL constant $0.55$. 
Thus the truncation error of the simulations can be related to the cell size alone. 
Since the discretization methods applied have first order precision, 
the error is proportional to the cell size squared. 
Thus, in the most refined refinement level, 
the order of magnitude of the truncation error is $10^{-6}$ for S1, S2 and S3 
and $10^{-4}$ for S5 and S6. 
The automatic mesh refinement guarantees that during the simulation the clouds 
are entirely inside the highest precision level.

\begin{figure*}[t]
  \includegraphics[width=0.90\linewidth]{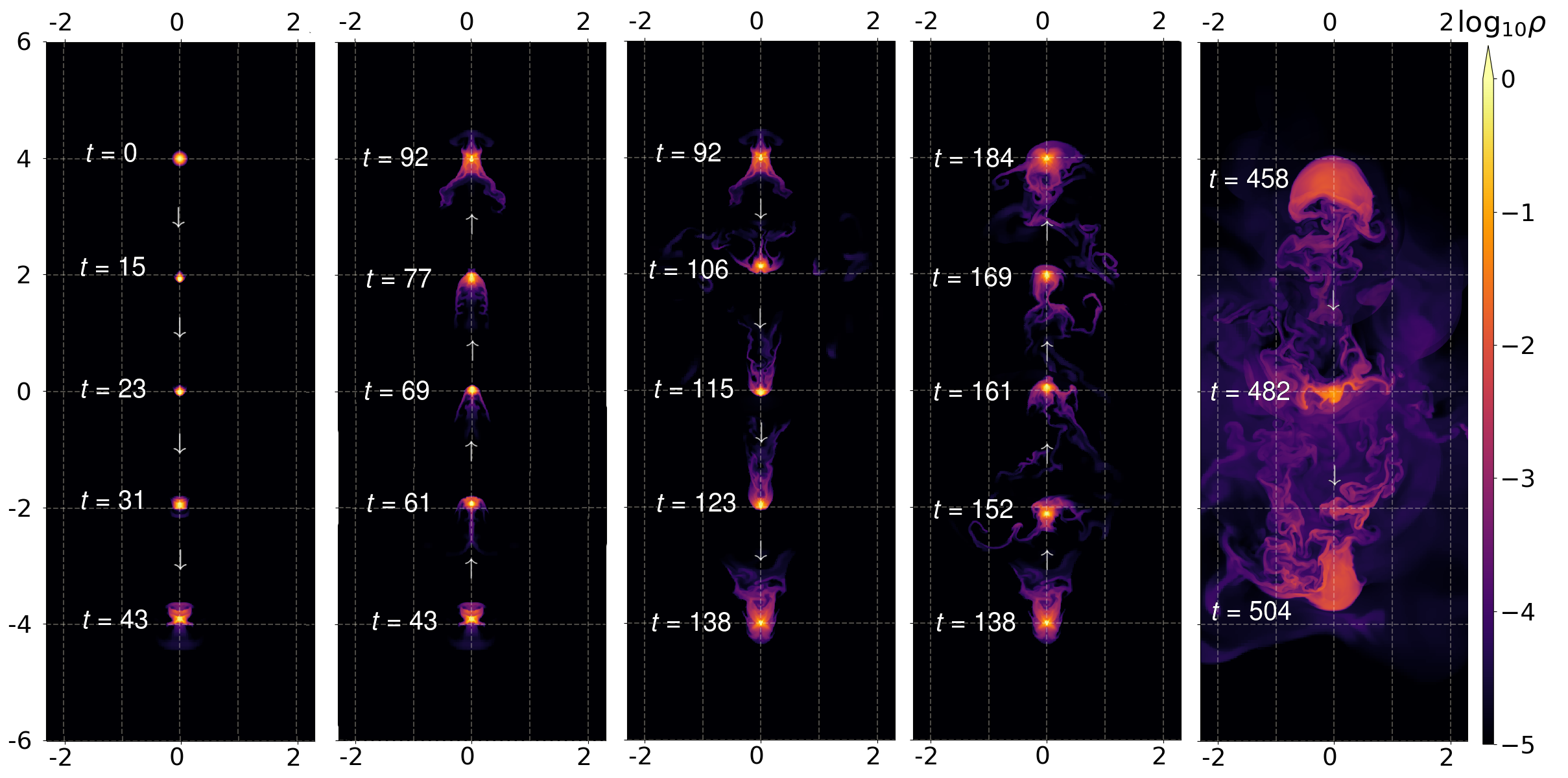}
     \caption{Simulation S1: 
Selected snapshots of the density (with logarithmic color coding)
for a head-on collision of the gas cloud with the BS.}
     \label{fig:cgt12geral}
\end{figure*}

\begin{figure}[t]
  \centering
\hspace{-0.3cm}{\includegraphics[width=0.9\columnwidth]{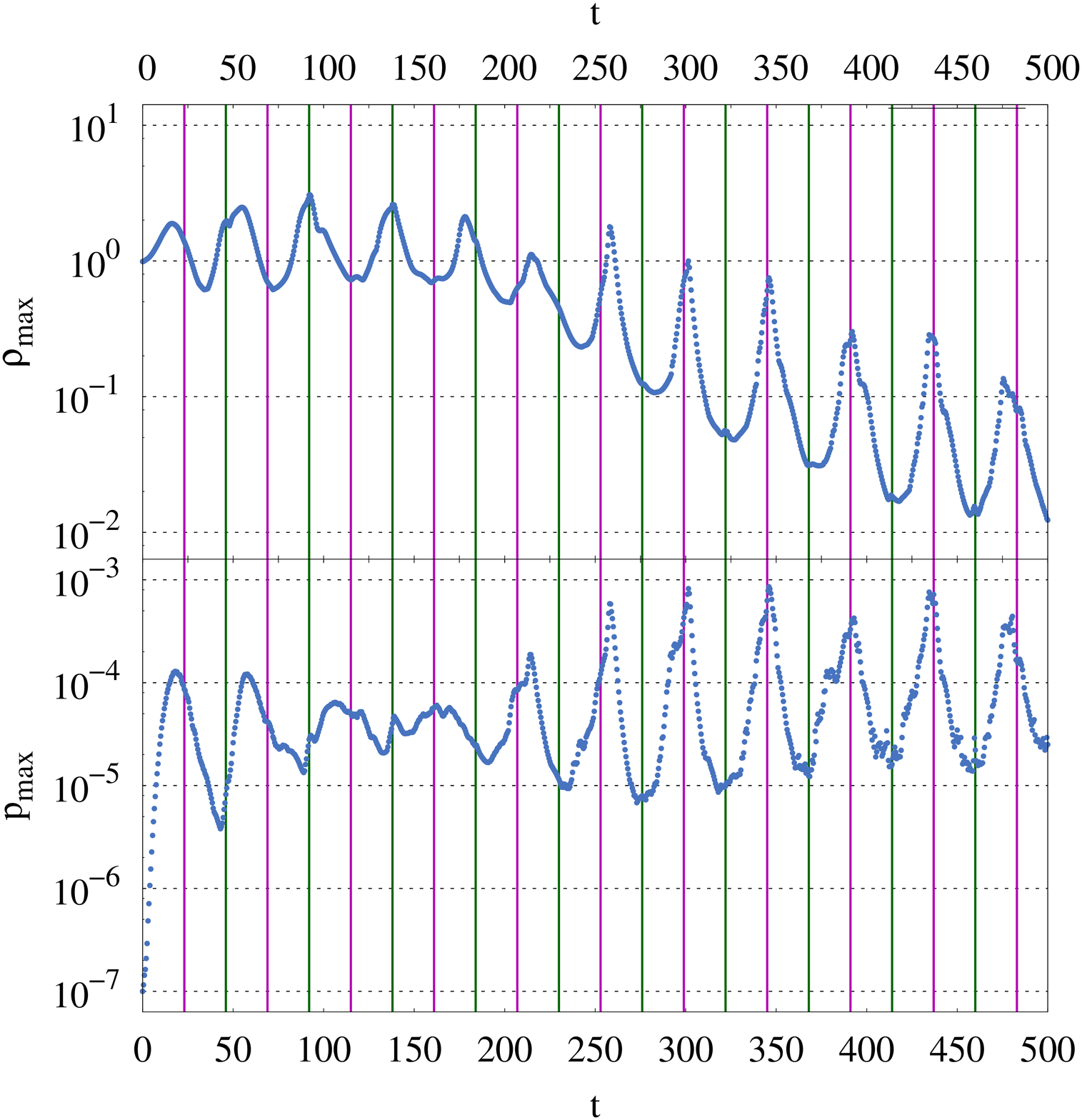}}
{\includegraphics[width=0.9\columnwidth]{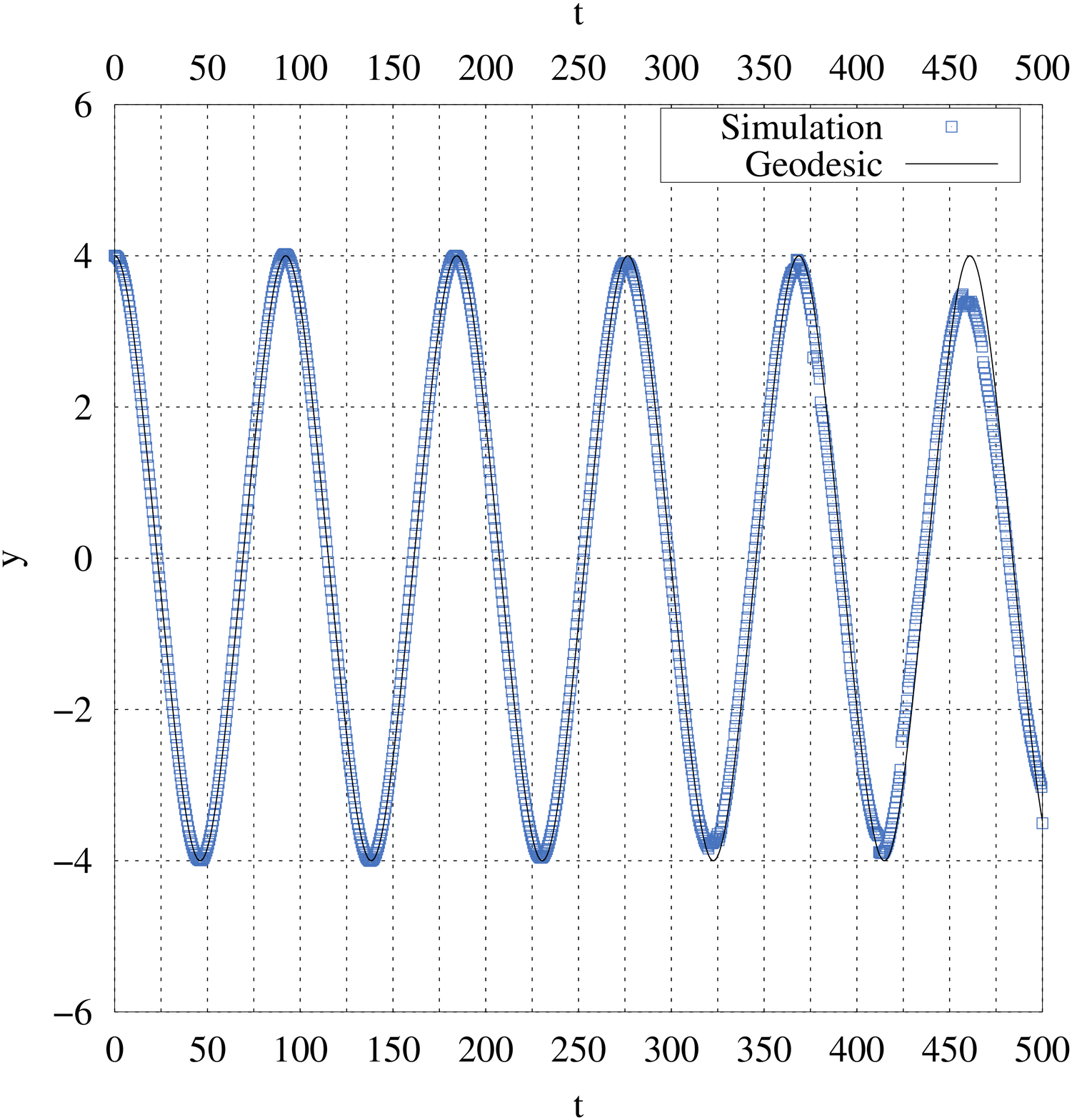}}
     \caption{Simulation S1:
(a - Upper panel) Maximum density $\rho_{\rm max}$
and maximum pressure $p_{\rm max}$ vs time $t$. 
The vertical lines highlight the pericenter (purple) and apocenter (green) passages of the corresponding test particle geodesic.
(b -  Lower panel) Position $(x=0,y)$ of the maximum density $\rho_{\rm max}$ 
of the cloud and the corresponding test particle geodesic (black)
vs time $t$.}
     \label{fig:cgt12rho_geo}
\end{figure}

\begin{figure}[t]
  \centering
\hspace{-0.3cm}{\includegraphics[width=0.9\columnwidth]{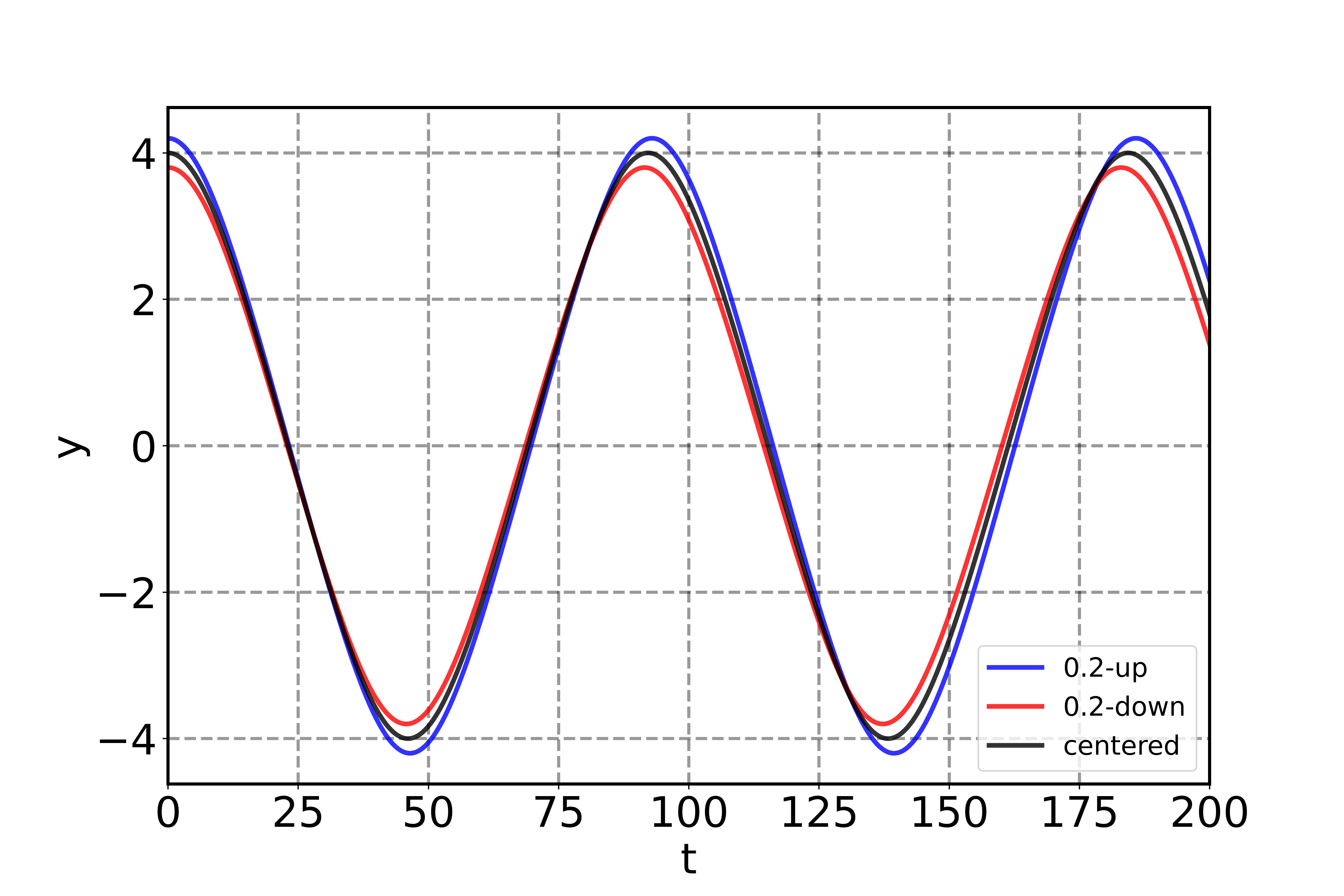}}
{\includegraphics[width=0.45\columnwidth]{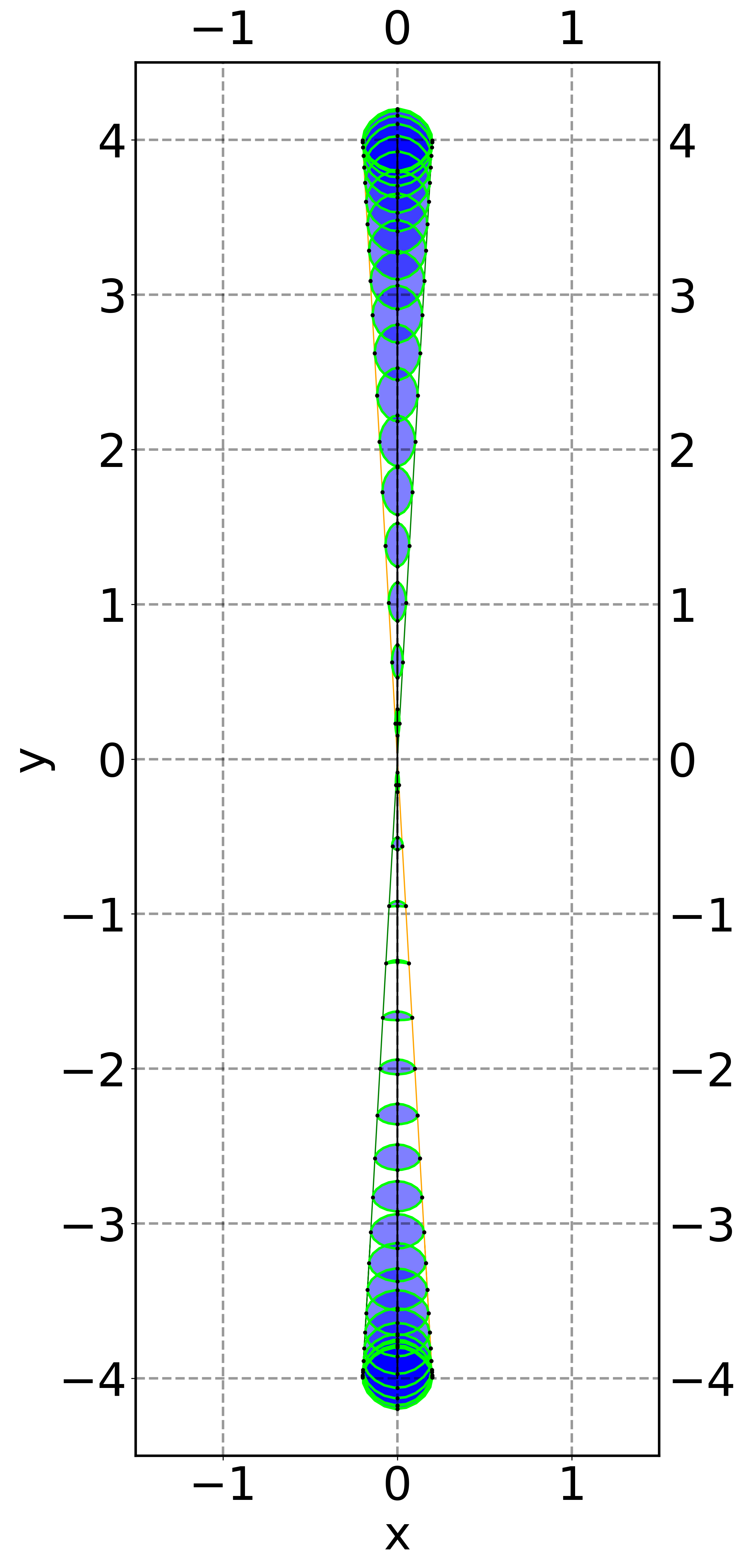}}
{\includegraphics[width=0.45\columnwidth]{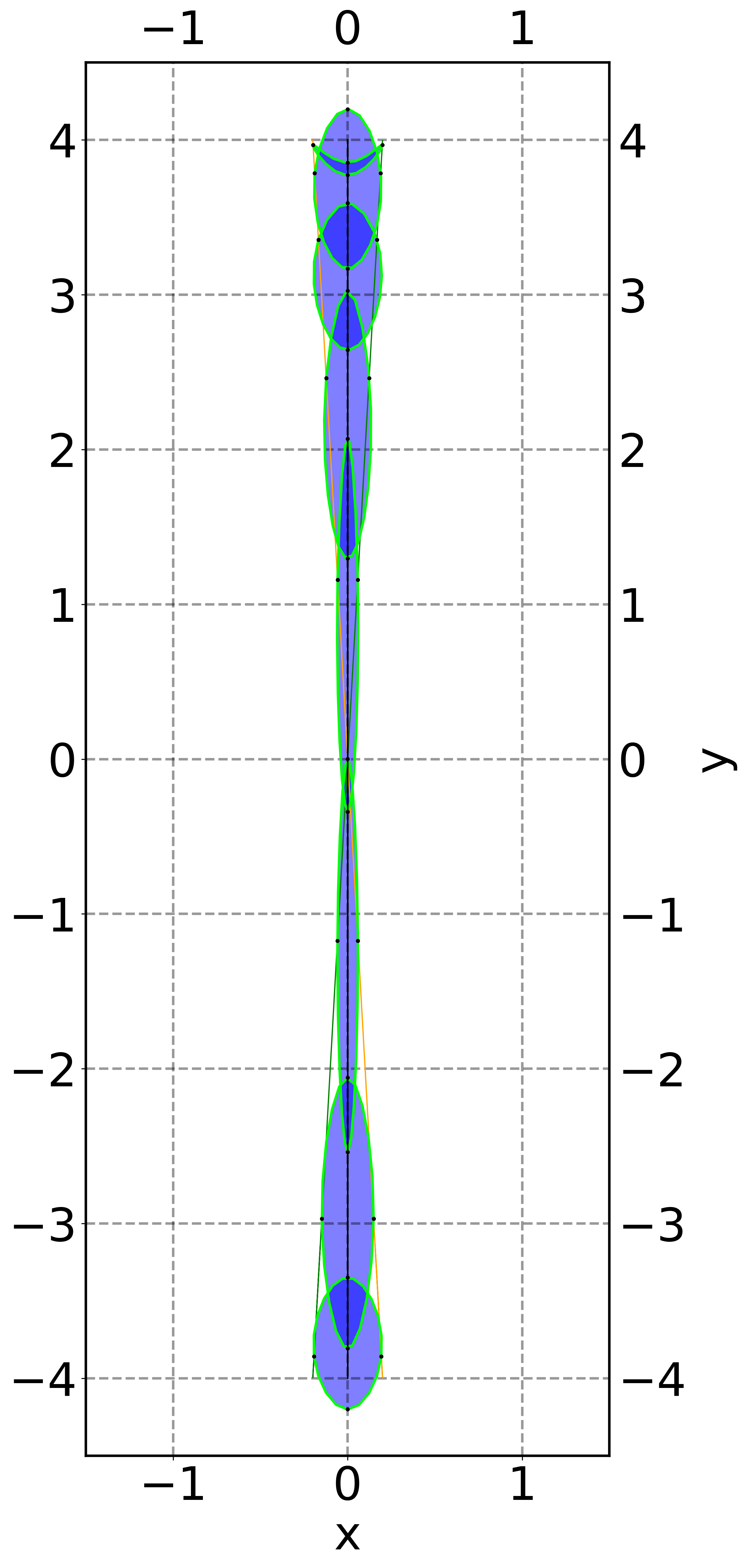}}
     \caption{Simulation S1: Comparison with geodesics of surface points of the cloud.
(a - Upper panel) Position $(x=0,y)$ of geodesics of upper edge, center and lower edge points vs time $t$.
(b,c -  Lower panels) Geodesic motion of cloud surface points during the first half period (left) and the last half period (right) of the cloud motion.
}
     \label{fig:geodesic_comp1}
\end{figure}

\section{Small and dense clouds (S1, S2 and S3)}

In this section we report the simulations regarding small ($R=0.03$) 
and dense ($\rho_{0}=1$) clouds nearby the BS,  
that are centered initially at $r_{0}=4$. 
The parameter choice for these simulations is aimed 
at approximating the test particle limit, 
at least in the beginning of the simulations. 
Starting slightly inside the radius $R_{\rm BS}$ of the BS, 
the simulations also represent a rather unique scenario, 
since analogous phenomena would not be feasible 
for compact objects endowed with a hard surface or an event horizon.

\subsection{S1 -- Head-on collision}

In the simulation S1 the cloud starts from rest ($v^r=v^\phi=0$)
with its center located at $(x=0,y=4)$.
Being gravitationally accelerated towards the BS center $(x=0,y=0)$,
the cloud then passes the center with velocity $v^{r}=0.74$.
As predicted by the geodesic motion of a test particle 
with the same initial conditions as those of the cloud,
the gas then decelerates, until it reaches the opposite 
of its starting position $(x=0,y=-4)$,
from where it moves back again towards the BS center. 
The tracking of the maximum density position,
exhibited in Fig.~\ref{fig:cgt12rho_geo}(b), 
shows consistency with the respective geodesic motion 
throughout the simulation. 
Although the maximum density position follows the geodesic, 
it can be seen on the snapshots 
selected in Fig.~\ref{fig:cgt12geral},
that during the motion of the cloud
debris is released from it, 
making the cloud lose its initial shape.

To illustrate the expected tidal effects,
we compare in Fig.~\ref{fig:geodesic_comp1} the geodesics of surface points of the cloud.
The upper panel shows the position $(x=0,y)$ of the geodesics of the upper edge point, the center and lower edge point versus time $t$, revealing slight changes in the periods that depend on the relative position.
In the lower panels we illustrate the geodesic motion of the points of the cloud surface (taken at the beginning of the run $t=0$) and following them during the first half period (left figure) and the last half period (right figure) of the cloud motion.The deformation of the initially spherical cloud surface indicates the squeezing effects found as a result of our simulations.


Studying now the cloud motion in more detail, we note that
the initial acceleration of the cloud 
is accompanied by a compression transverse to the gas motion, 
that is related to the medium's resistance to the motion. 
The change of the direction of motion together with thermal rebound 
causes several of such compressions which are dominant until $t=200$. 
After this point, another type of compression-expansion 
oscillation takes place. 
Namely, the now more extended cloud is more susceptible to tidal forces, 
which are maximal at the BS center and minimal an the orbit's apocenters. 
These forces then give rise to compression-expansion cycles 
endowed with half of the period of the orbit, 
where maximal compression is found
when the cloud passes through the origin (pericenter) 
and maximal expansion at the apocenters of the orbit. 
The cloud then continually increases in size at the apocenters, 
providing feedback to the tidal compression at the center,
making the cloud even broader.
These self-sustained processes then dismantle the cloud. 
The cycles can be tracked by following the maximum density 
and the maximum pressure of the gas,
as shown in Fig.~\ref{fig:cgt12rho_geo}(a).

Another important debris formation mechanism occurs 
by the formation of short-term double tails,
which lose their shape through gas-tail collision,
when the motion of the cloud changes direction.
In the collision the gas of the cloud leads to shock waves in the tails,
while the tails also create minor shock waves,
that travel and bounce inside the cloud. 
Also, low density debris constantly hits the cloud, 
since its trajectory does not necessarily follows the geodesic. 
A combination of these two types of gas-gas interaction then
triggers turbulence in the cloud's tail and surface, 
extracting chunks of fluid.

The combination of these two mechanisms 
will eventually destroy the original shape of the cloud 
as seen in the last snapshot of Fig.~\ref{fig:cgt12geral}. 

\subsection{S2 -- Closed elliptic orbit}

\begin{figure*}[ht]
  \includegraphics[width=0.90\linewidth]{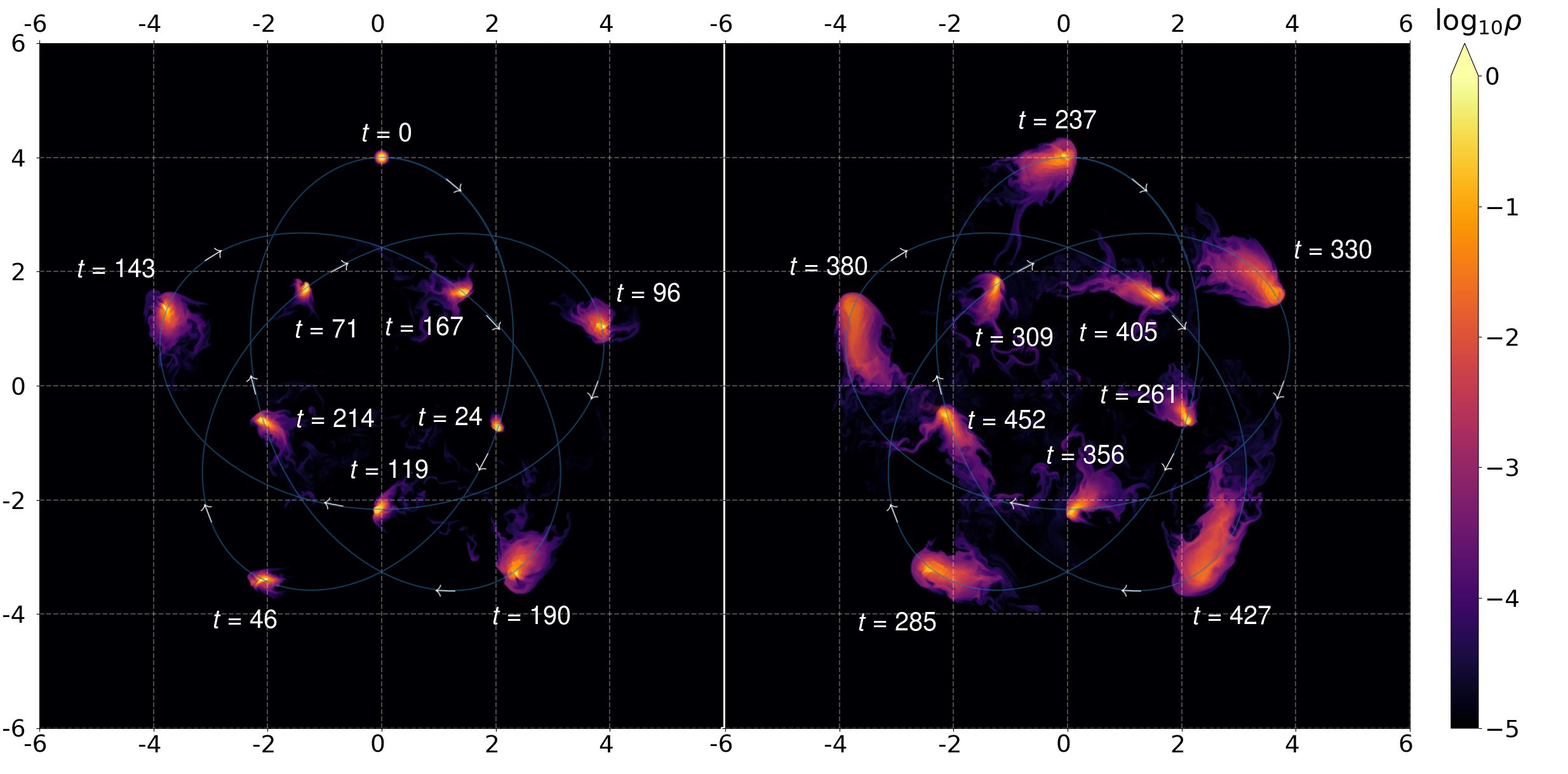}
     \caption{Simulation S2:
Selected snapshots of the density (with logarithmic color coding)
for an elliptic orbit of the gas cloud around the BS.}
     \label{fig:cgt15snaps}
\end{figure*}

\begin{figure}[t!]
 \centering
{  \includegraphics[width=0.9\columnwidth]{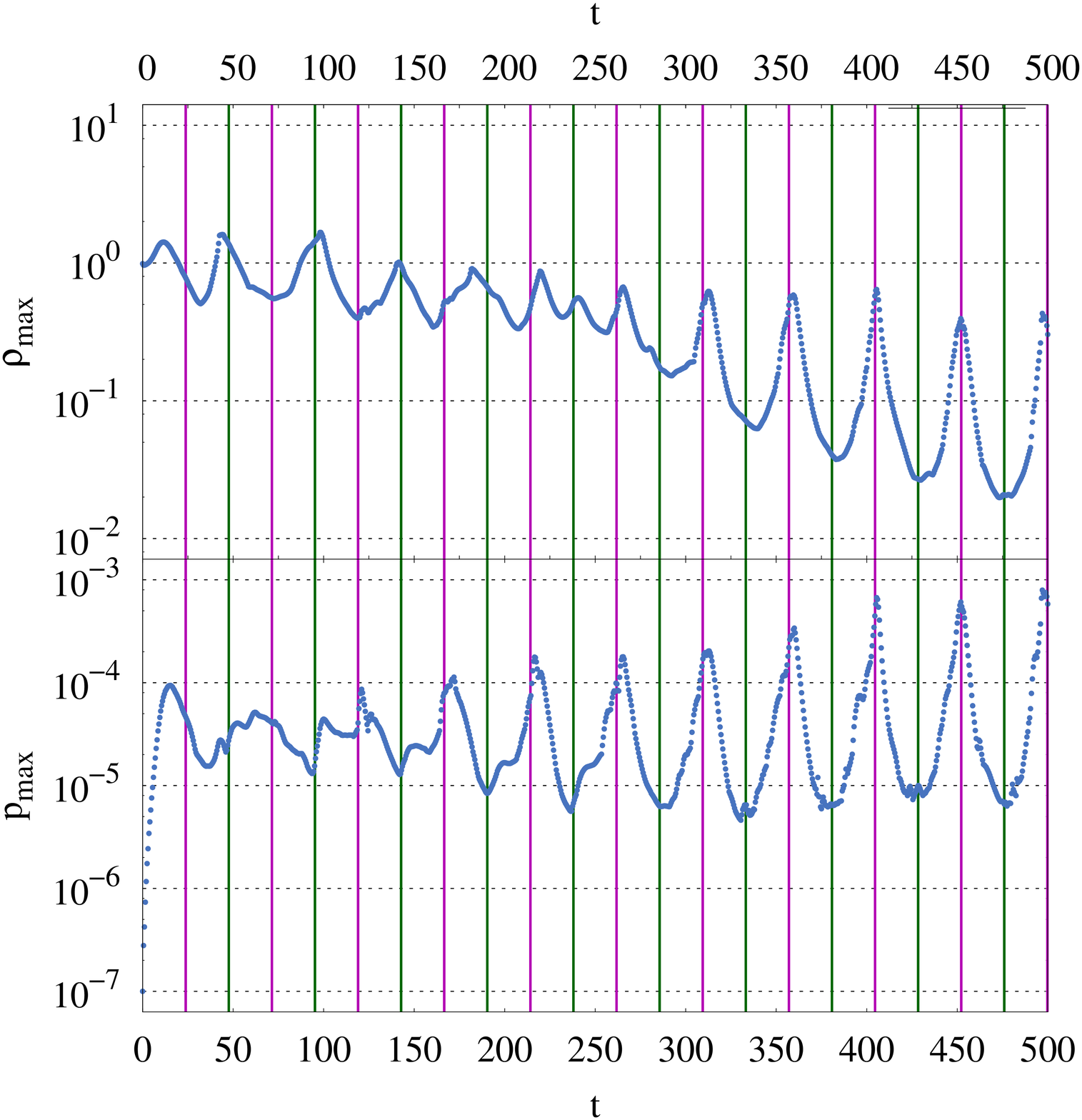}}
{  \includegraphics[width=0.9\columnwidth]{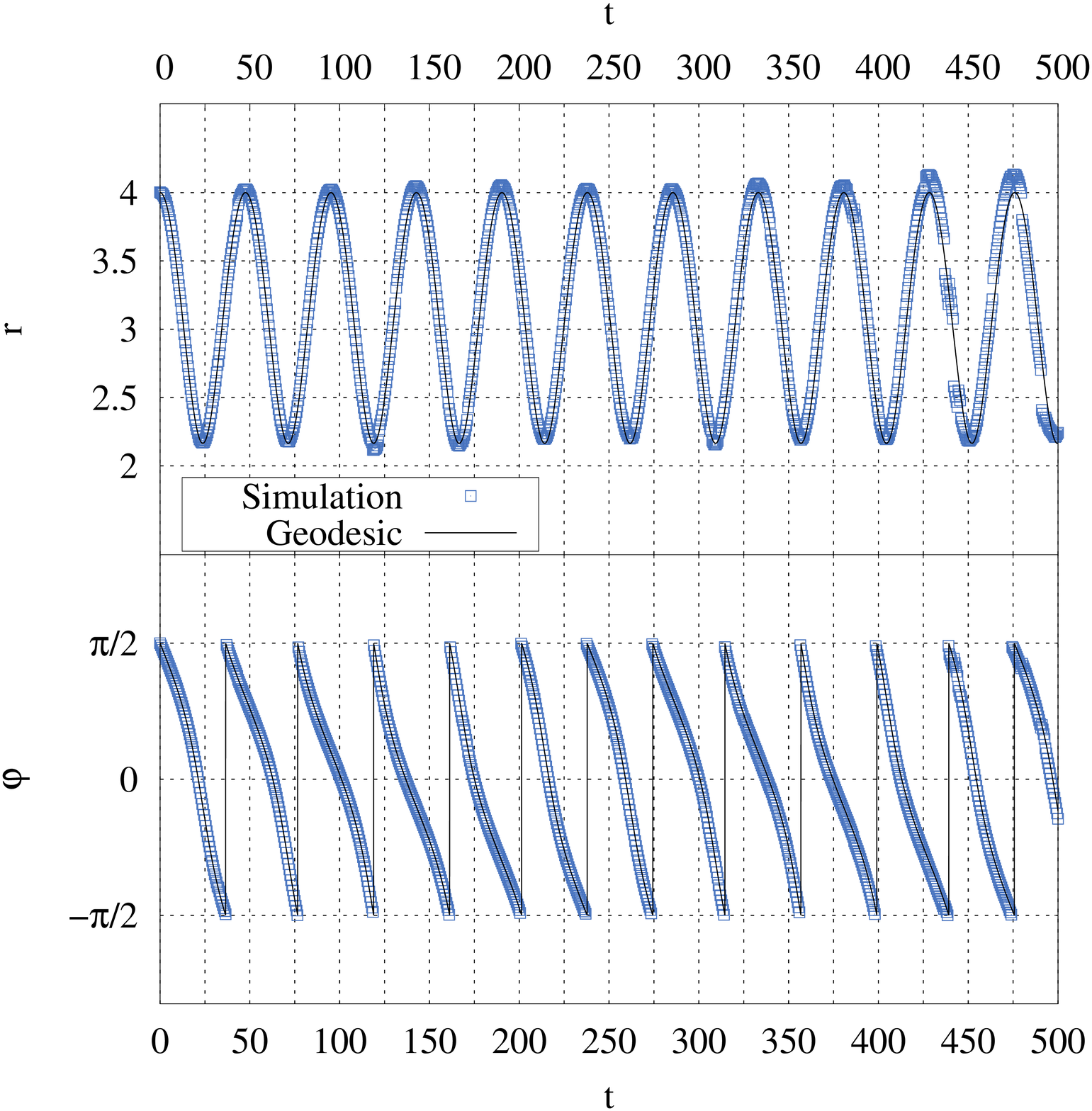}}
\caption{Simulation S2:
(a -Upper panel) Maximum density $\rho_{\rm max}$
and maximum pressure $p_{\rm max}$ vs time $t$. 
The vertical lines highlight the pericenter (purple) and apocenter (green) passages of the corresponding test particle geodesic.
(b - Lower panel) Position $(r,\varphi)$ of the maximum density $\rho_{\rm max}$ (blue)
of the cloud and the corresponding test particle geodesic (black)
vs time $t$.}
     \label{fig:cgt15rho_geo}
\end{figure}

The second simulation reported here is done by matching 
the initial conditions of the cloud with those 
of a test particle on a closed elliptic orbit.
It is obtained by
giving the gas an angular momentum $\mathbf{L}=1.7890$,
while all the other simulation parameters remain the same. 
A selection of snapshots from S2 is shown 
in Fig.~\ref{fig:cgt15snaps}(a) and (b).
Once again, the position of the maximum density of the cloud 
follows the geodesic motion, as can be seen 
in Fig.~\ref{fig:cgt15rho_geo}(b), 
where the radial distance $r$ and azimuthal angle $\varphi$ 
of $\rho_{\rm max}$ are shown and compared to the
corresponding geodesic motion.

Now the angular motion of the cloud prevents 
the previous abrupt change of direction in its trajectory. 
Thus no significant cloud-tail interaction is observed. 
On the other hand, the rarefied gas 
(with a density on the order of magnitude of the atmospheric density) 
that is extracted from the cloud since the beginning of the simulation, 
falls almost freely towards the BS center. 
Although not significant for the cloud itself in terms of fluid-loss, 
such tiny debris is accelerated, reaching high velocities 
(similar to the ones observed in S1) and, 
after crossing the BS center it re-encounters the cloud 
at about $t=50-60$. 
Although no changes in the course of the cloud center are found 
as a result of such an encounter, 
the collision of the outgoing debris and the cloud, 
which is at that moment traveling towards the center, 
generate turbulence through a shock wave encounter. 
Indeed, shock waves formed by rarefied gas fronts moving outward 
keep hitting the cloud and its tail during the simulation, 
making the tail and the cloud's border turbulent.   

On the other hand, cycles of compression-expansion are still present. 
The first peak of the maximum pressure,
shown in Fig.~\ref{fig:cgt15rho_geo}(a),
is related to the initial conditions of the cloud. 
The start of the motion of the cloud through the atmosphere, 
which must indeed make its way through the medium, 
causes the first compression of the gas. 
This compression then thermally rebounds and starts to oscillate. 
This mode of oscillation competes with the cycles 
of expansion-compression similar to the ones 
found in the second half of the simulation S1. 
Since the latter increase in amplitude, 
after $t=200$ these modes dominate the simulation. 
Indeed, now synchronized with the apo- and pericenters, 
these cycles become the main reason for the cloud's deformation.
The finale of the simulation, 
illustrated by the snapshots in Fig.~\ref{fig:cgt15snaps}(b), 
shows how the increasing amplitude of these cycles 
increasingly deforms the broadened and elongated cloud and thus, 
similarly to S1, turbulence is triggered at its borders
and extracts gas.

\begin{figure}[t!]
 \centering
{  \includegraphics[width=0.9\columnwidth]{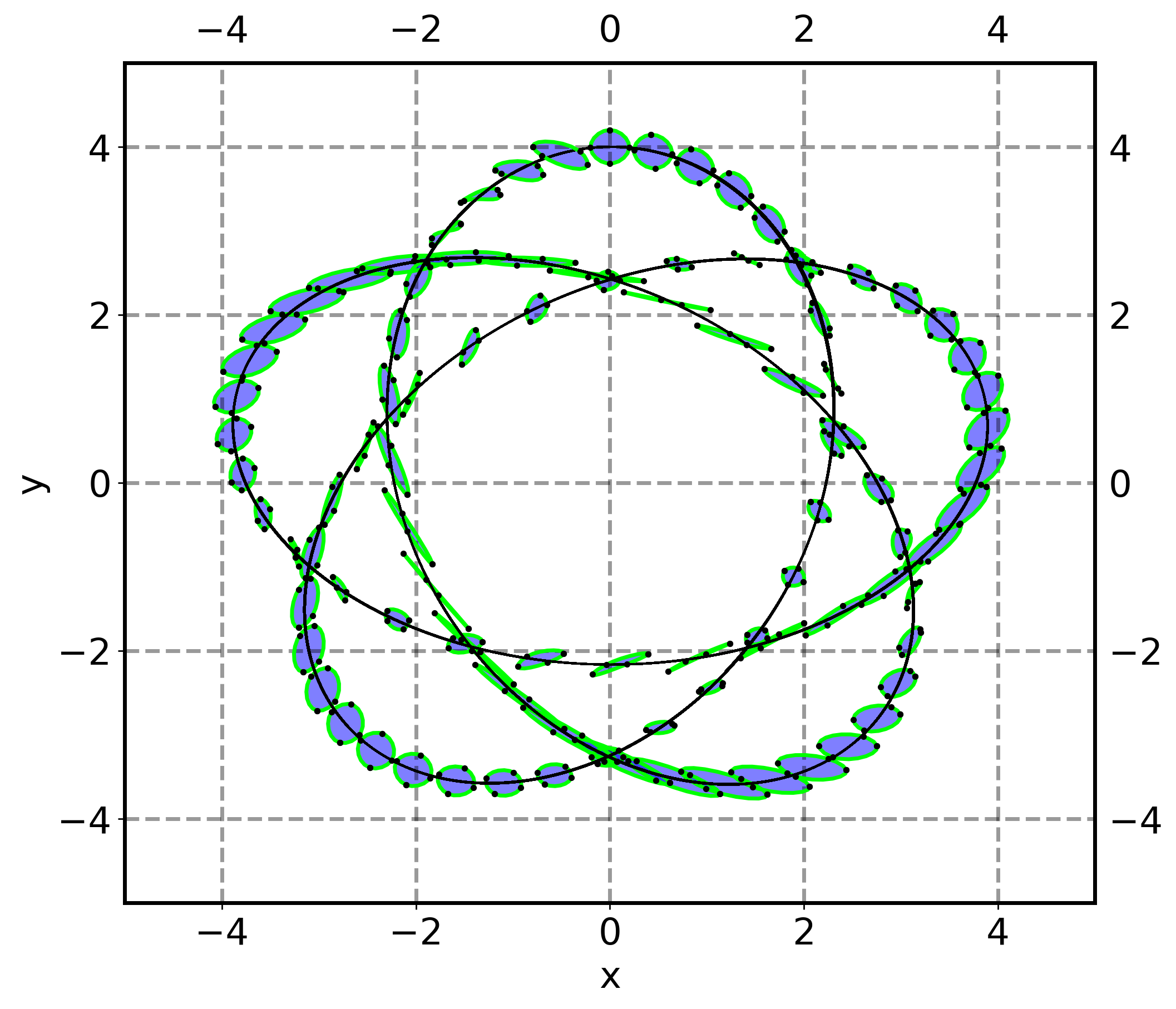}}
{  \includegraphics[width=0.9\columnwidth]{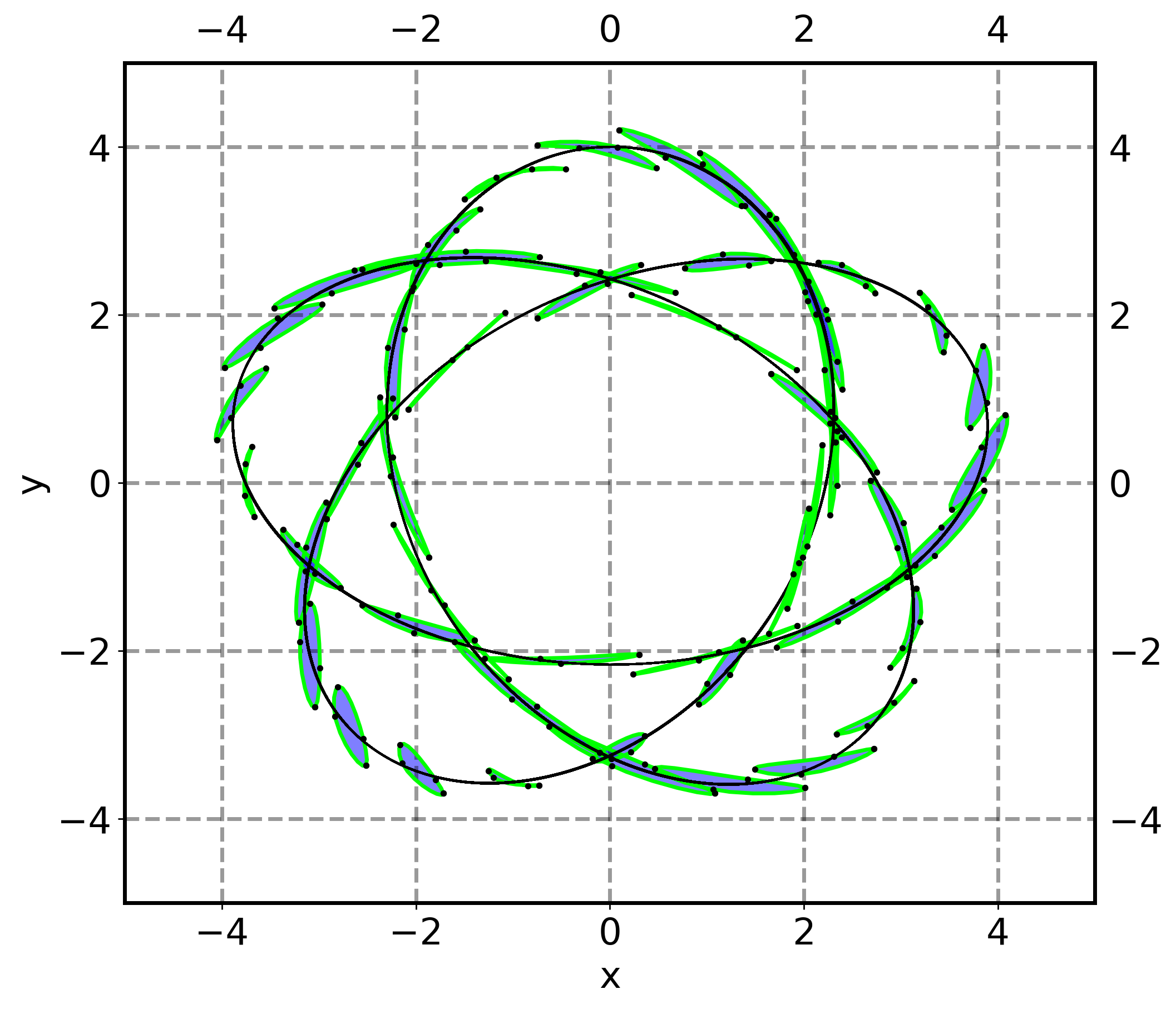}}
\caption{Simulation S2: Comparison with geodesics of surface points of the cloud.
(a - Upper panel) Geodesic motion of cloud surface points during the first half of the cloud motion.
(b - Lower panel) Geodesic motion of cloud surface points during the second half of the cloud motion.
}
     \label{fig:geodesics_comp2}
\end{figure}

We illustrate the purely tidal effects,
in Fig.~\ref{fig:geodesics_comp2}, where we evolve the geodesics of surface points of the cloud.
The upper panel shows the geodesic motion of the points of the cloud surface (taken at the beginning of the run $t=0$) during the first half of the run, while the lower panel illustrates the second half. Again, the deformation of the initially spherical cloud surface due to tidal effects can already be expected from the geodesics deviations. Although significant, these deviations do not take into account hydrodynamic effects that also influence the shape of the cloud.

\subsection{S3 -- Circular orbit}

\begin{figure*}[t!]
  \includegraphics[width=0.90\linewidth]{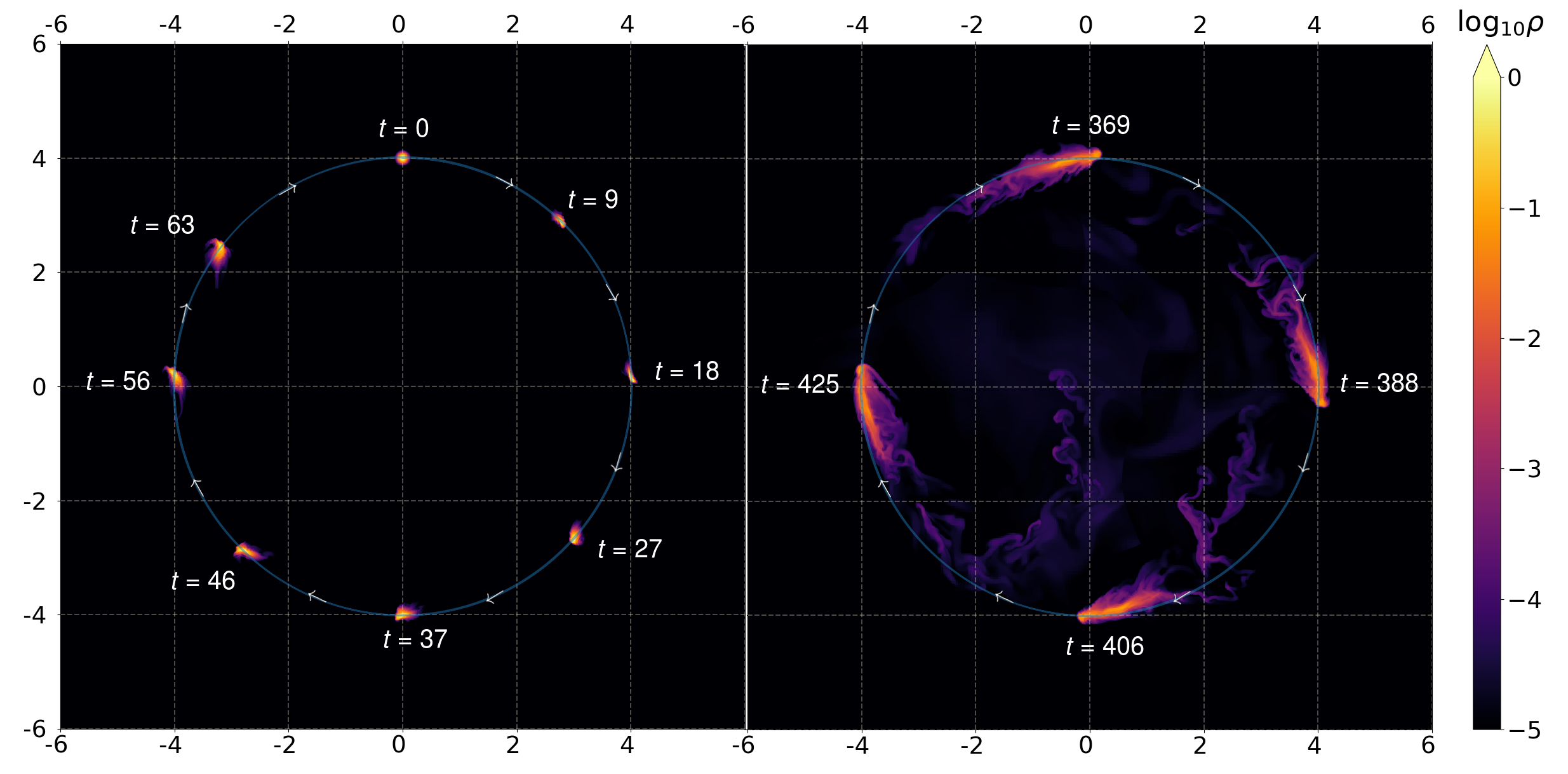}
     \caption{Simulation S3:
Selected snapshots of the density (with logarithmic color coding)
for a circular orbit of the gas cloud around the BS.}
     \label{fig:cgt17snaps}
\end{figure*}

\begin{figure}[t!]
\centering
{  \includegraphics[width=0.9\columnwidth]{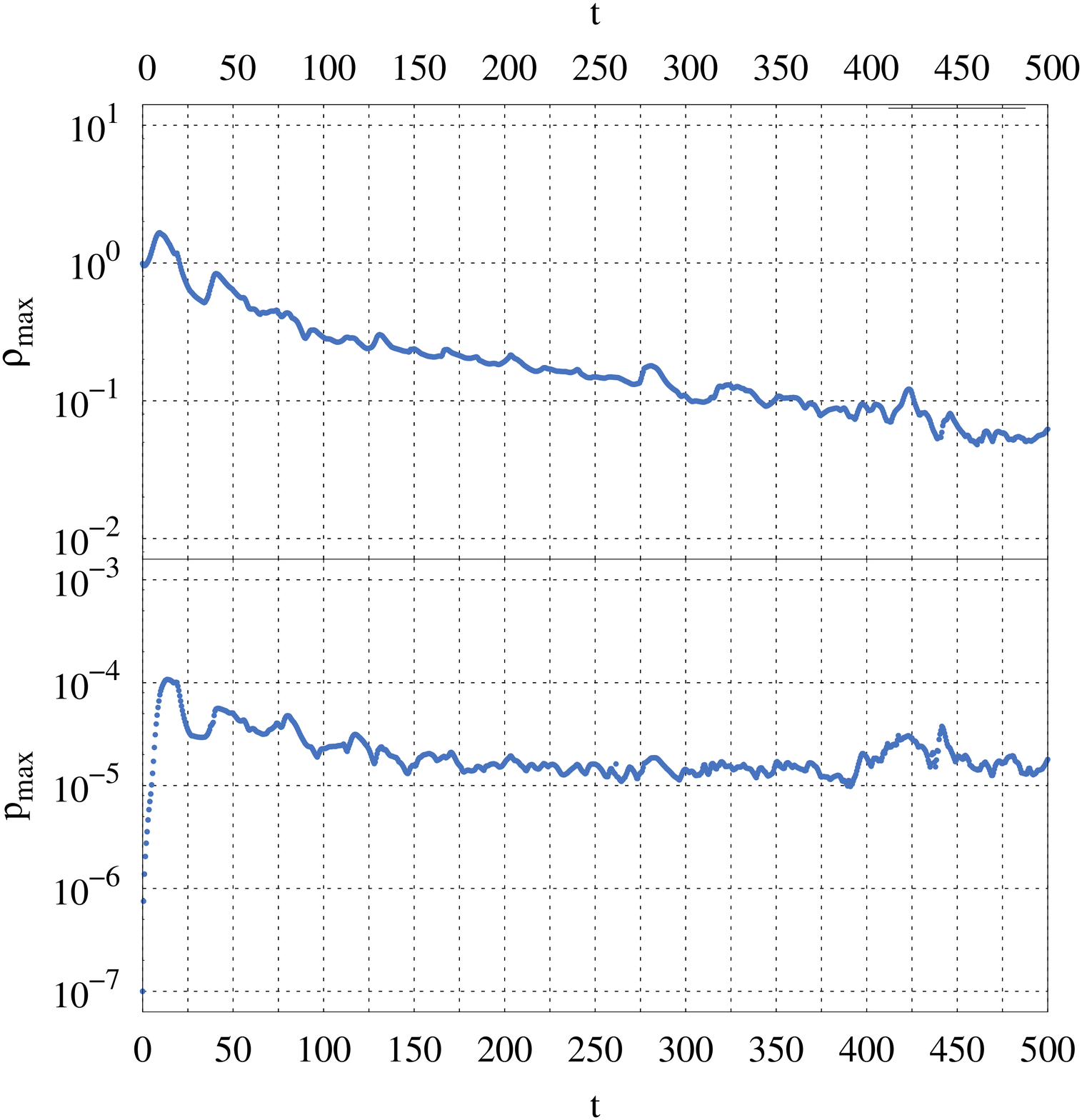}}
{  \includegraphics[width=0.9\columnwidth]{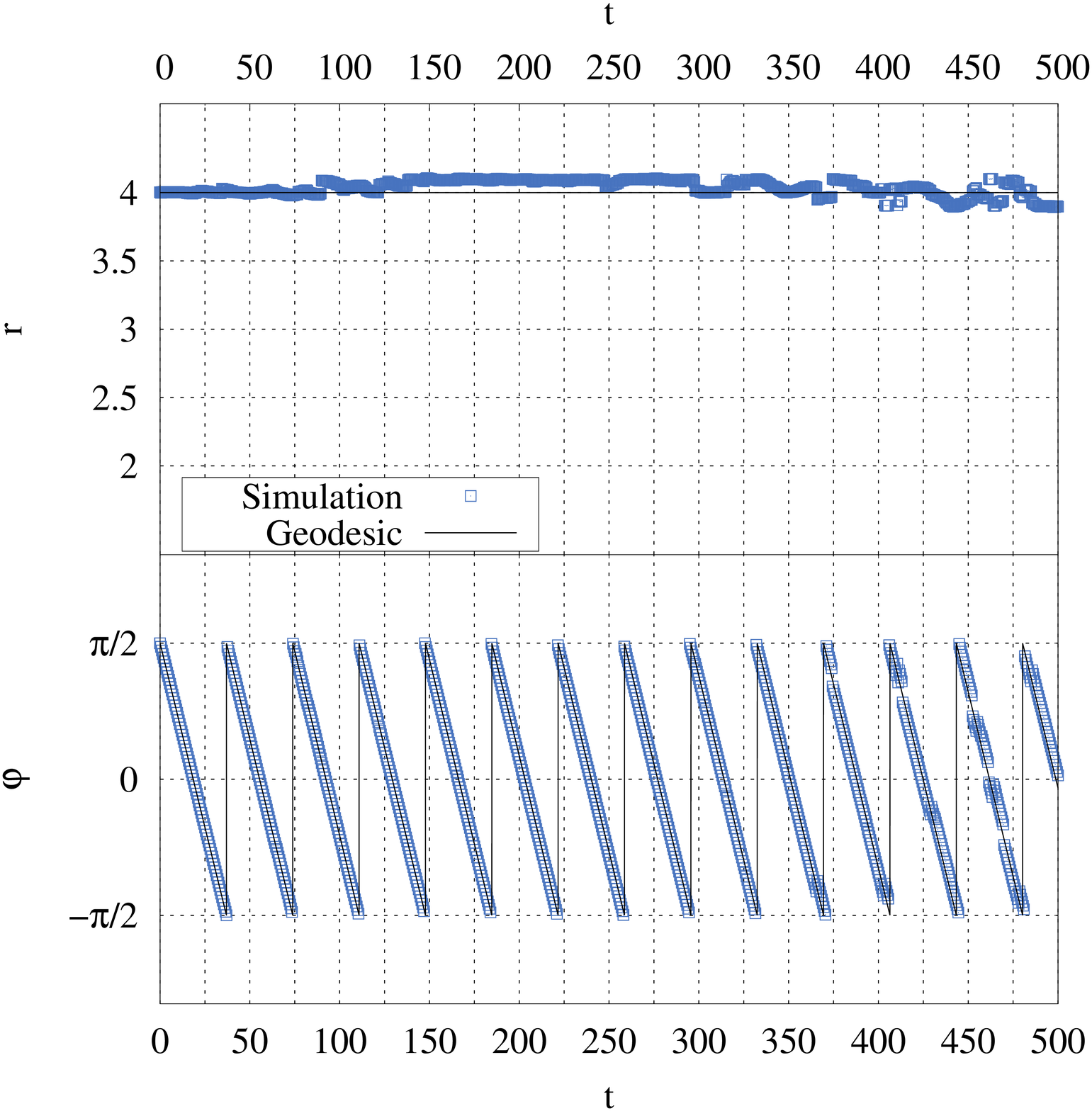}}
     \caption{Simulation S3:
(a - Upper panel) Maximum density $\rho_{\rm max}$
and maximum pressure $p_{\rm max}$ vs time $t$.
(b - Lower panel) Position $(r,\varphi)$ of the maximum density $\rho_{\rm max}$ (blue)
of the cloud and the corresponding test particle geodesic (black)
vs time $t$.}
     \label{fig:cgt17rho_geo}
\end{figure}

Finally, we have performed a third simulation, S3, regarding 
close-by compact clouds. 
The angular momentum of the cloud was set to $\mathbf{L}=3.583$ for S3. 
This angular momentum corresponds to the one of a circular orbit 
of a test particle at coordinate radius $r=4$ around the BS. 
A set of selected snapshots for this simulation is shown
in Fig.~\ref{fig:cgt17snaps}.
The position of the maximum density $\rho_{\rm max}$
follows again the geodesic motion, 
as seen in Fig.~\ref{fig:cgt17rho_geo}.
   
For circular orbits there are no peri- or apocenters, 
meaning that the cycles of contraction-expansion of the gas 
should not be present, since the tidal forces do not change
for a constant radius. 
Indeed, as seen in Fig.~\ref{fig:cgt17rho_geo}(a), 
the maximum density $\rho_{\rm max}$ 
and the maximum pressure $p_{\rm max}$
do not feature strong periodic oscillations. 
The first peak of the pressure is due to the initialization 
of the cloud movement, and the decreasing value of the maximum density 
is related to the tail formation. 
Indeed, through the course of the simulation a prominent tail is formed. 
Tail formation is now mainly related 
to the cloud movement through the medium, 
and since the cloud does not suffer transverse expansion, 
the tail is free to increase in the direction parallel to the gas motion. 
The elongated tail is then also susceptible 
to the Kelvin-Helmholtz instability 
and debris-cloud collisions similar to the ones found in S1 and S2.
Thus gas is slowly extracted from the cloud-tail 
formed structure through turbulence. 
Cloud-debris collision cannot be prevented in the circular orbit,
since rarefied debris generation is also observed for S3.

\subsection{Discussion}
 
Comparing these results with the near-by clouds 
simulated in \cite{Meliani:2017ktw}, 
it is seen that the different BS model, a rotating mini-BS,
employed in \cite{Meliani:2017ktw}
and the different initial conditions for the cloud 
create contrasting gas behavior. 
Regarding the head-on collision, for a spherically symmetric BS, 
instead of Lense-Thirring dominated flows as present for rotating BSs,
local tidal forces dominate the fluid. 
For that reason effects such as compression/expansion cycles 
combined with gas-tail and gas-debris collisions 
deform the cloud much faster than in the presence of a rotating BS. 
By giving the cloud angular momentum, 
such effects become less dramatic,
since the trajectory of the fluid in the radial direction 
becomes less substantial. 
Indeed, by choosing a circular orbit, 
only gas-debris collision remains 
as a gravity caused disruption mechanism. 
On the other hand, for this circular orbit 
the gas must be provided with initial angular momentum.
Then its movement through a static medium generates a prominent tail.  

The debris formation mechanisms discussed in this section 
are summarized below, and their importance for each simulation 
is indicated in table \ref{tab:table2}.
\begin{enumerate}
    \item {\it Tail-cloud collision - } Present when dramatic changes 
in the direction of motion of the cloud are realized. 
The cloud's tail collides directly with the cloud's main body, 
eliminating gas from its surface through strong shear forces.
     \item {\it Compression-expansion cycles - } Present when apocenters 
and pericenters are encountered in the orbit. 
The cloud suffers increasingly periodic compression 
and subsequent expansion cycles, 
synchronized with the orbit's extrema. 
Such cycles deform the cloud, increasing its size 
and extracting gas from it during the contraction phases.
     \item {\it Debris-cloud collision - } Present when rarefied gas 
from the cloud's surroundings re-encounters the cloud 
after passage through the BS center. 
Such collisions trigger turbulence in the cloud's surface, extracting gas. 
\end{enumerate}

\begin{table*}[t!]
 {\footnotesize
    \begin{tabular}{|c|c|c|c|c|c} 
      \hline
     \textbf{Label} &  \textbf{Tail-cloud collision} &  \textbf{Compression-expansion cycles} & \textbf{Debris-cloud collision} \\
      \hline
      S1 & strong & strong & yes  \\
      \hline
      S2 & weak & middle & yes  \\
      \hline
      S3 & not present & not present & yes  \\
      \hline
    \end{tabular}
    }
    \caption{Debris formation mechanisms for near-by dense small clouds orbiting the BS.}
        \label{tab:table2}
\end{table*}

Based on these three simulations we put forward a conjecture
on the lifetime of small dense clouds nearby the 
 compact spherically symmetric BSs. 
Since for head-on collisions all three debris mechanisms are present, 
clouds on such trajectories seem to be the least stable ones
under the given conditions.
Less strong compression and expansion cycles 
as well as the absence of cloud-tail interaction 
will provide a longer lifetime for clouds on elliptical orbits.
Finally, by eliminating the dramatic changes of the tidal forces 
on the clouds, circular orbits will be the most stable ones. 

\section{Circular orbits of extended clouds (S4 and S5)}

In this section we report simulations of extended gas clouds 
on (initially) circular orbits around 
a compact spherically symmetric BS (S4),
and compare with the simulations of such clouds
around a Schwarzschild BH of the same mass (S5). 

\subsection{Setup}

We have chosen the setup 
to avoid the expansion-contraction cycles and tail-gas interactions,
that arise otherwise as shown in section 4. 
The angular momentum of test particles on circular orbits is given by
\begin{equation}
\mathbf{L}^2=\frac{g_{tt}^\prime}{g_{tt}}\Bigg(\frac{2}{r_0^3}-\frac{g_{tt}^\prime}{g_{tt}r_0^2} \Bigg)^{-1}.
\label{eq1}
\end{equation}
The initial radial position of the cloud center is chosen to be $r_0=10$,
a value of the radial coordinate where the metrics of the BS and the
corresponding Schwarzschild BH are already very similar,
as seen in Fig.~\ref{mxr_bs}(b).
Therefore the initial angular momentum required 
for the circular geodesics in these spacetimes is very similar as well,
$\mathbf{L}_{\rm BH}-\mathbf{L}_{\rm BS}\approx10^{-4}$.
On the other hand this value of the radial coordinate
is not too far from the inner spacetime regions,
where the metrics start to differ significantly,
representing therefore a meaningful choice for the comparison 
of the simulations.

The relatively big extension of the cloud, 
with standard deviation $R=0.3$, 
is chosen to achieve pressure gradients,
that are sufficient to realize considerable tidal effects.
These should substantially trigger accretion, but in a manner 
that the cloud could remain a reservoir of gas 
during the first stage of the simulation (before complete disruption). 
To realize the initially circular orbits we have chosen
in both simulations an angular momentum of $\mathbf{L}=6.075$
for the entire cloud, as obtained from eq.~(\ref{eq1})
for $r_0=10$.

\subsection{Simulations}

\begin{figure*}[p!]
  \includegraphics[clip,width=0.8\linewidth]{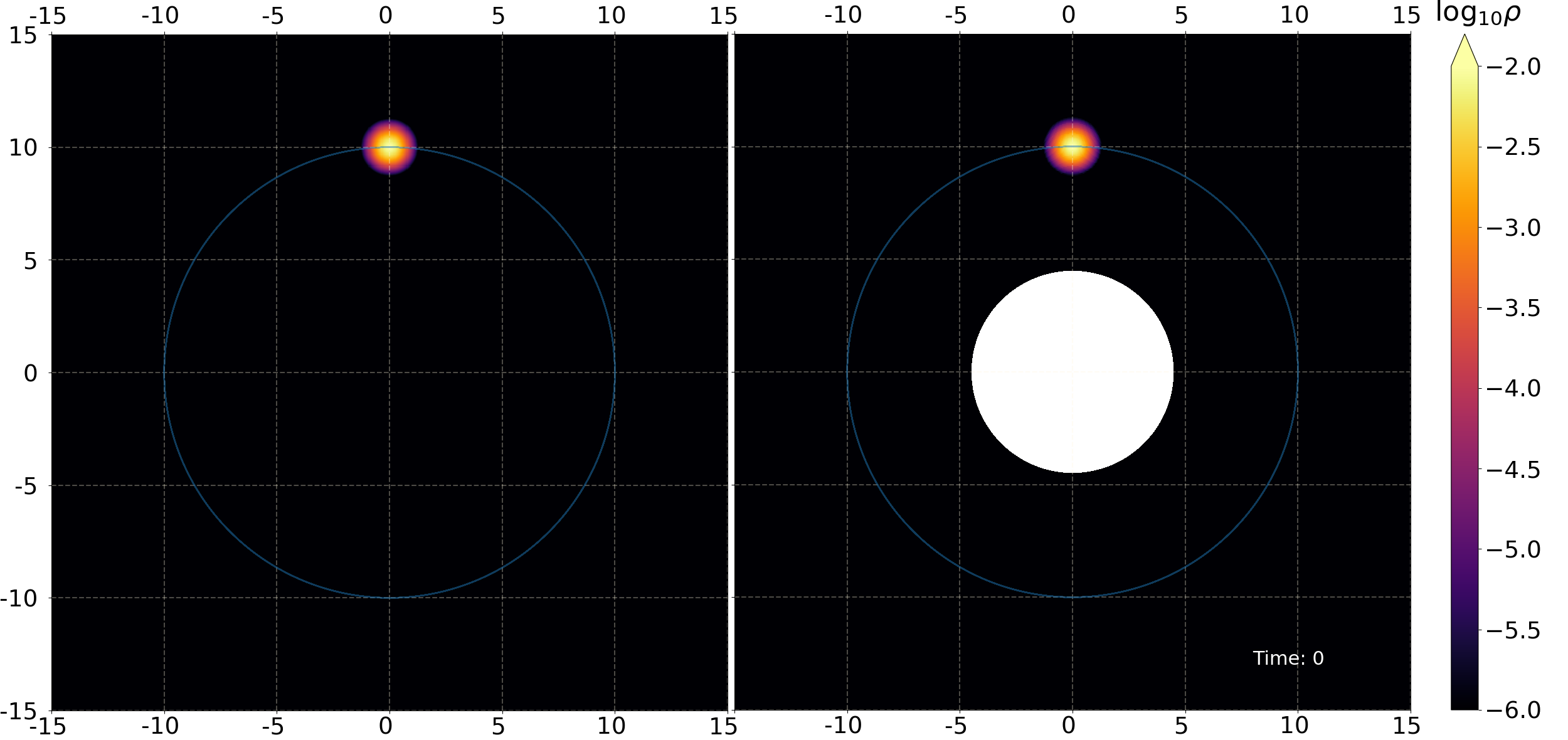}
  \includegraphics[clip,width=0.8\linewidth]{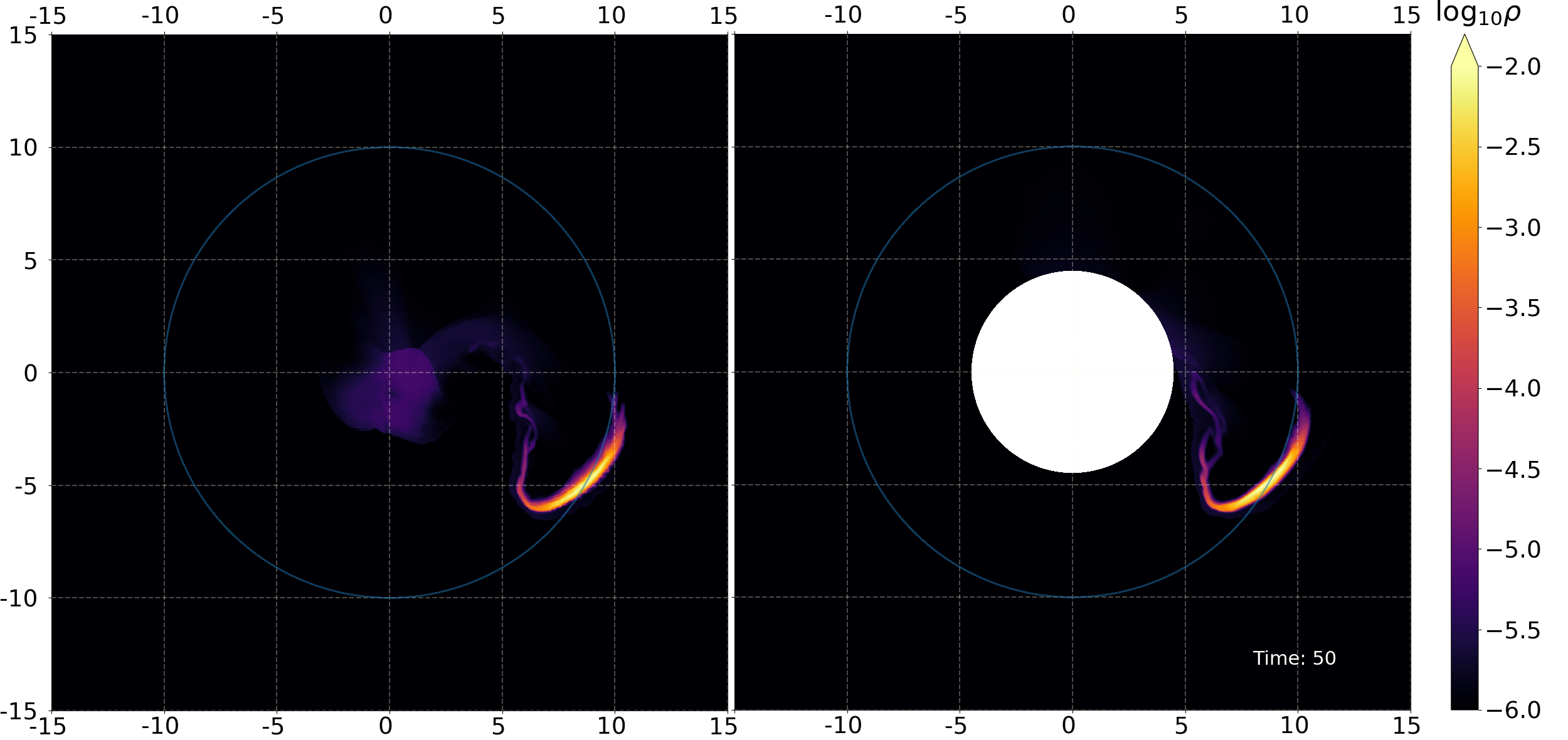}
  \includegraphics[clip,width=0.8\linewidth]{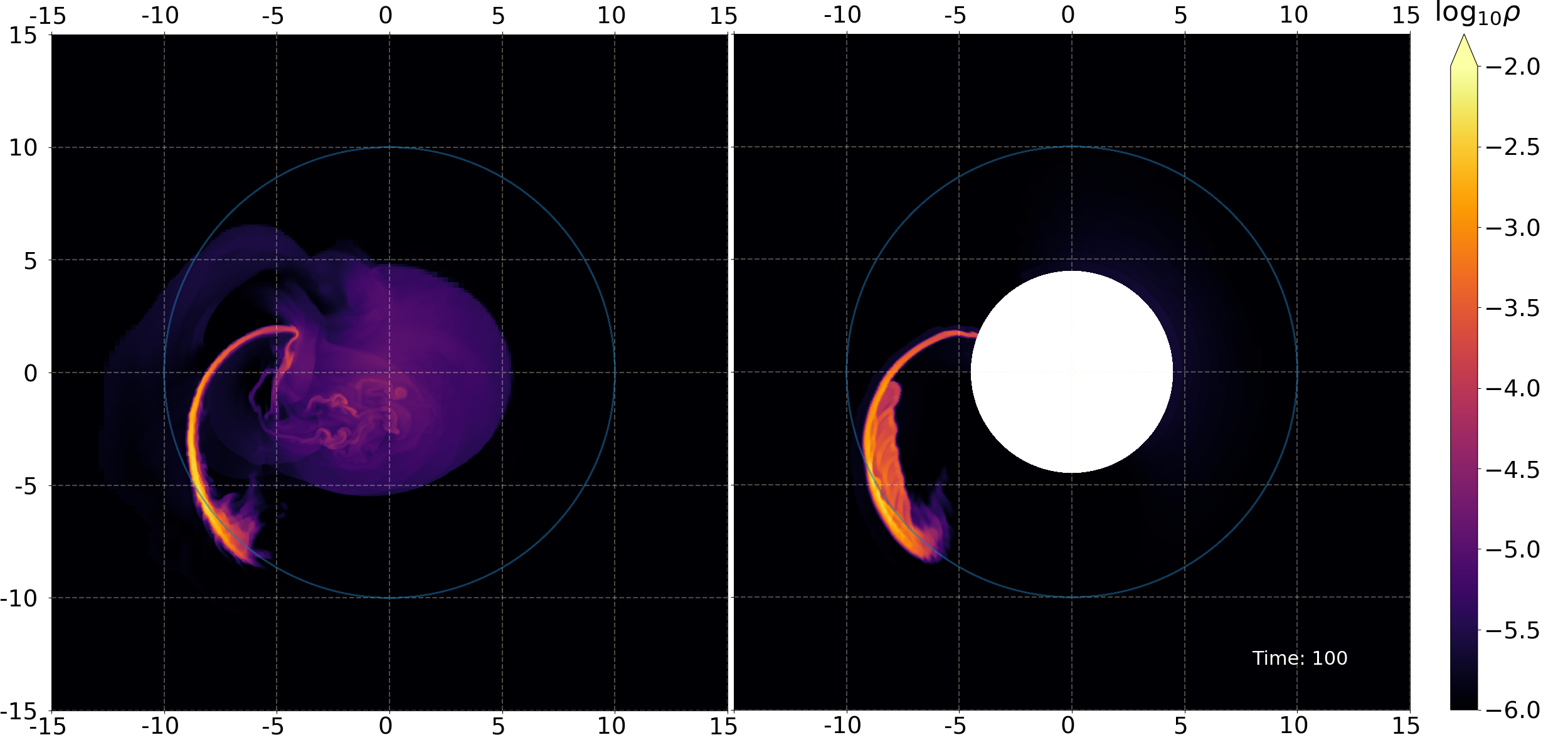}%
     \caption{Simulations S4 and S5:
Selected snapshots at time $t=0$, $50$ and $100$
of the density (with logarithmic color coding)
for an (initially) circular orbit 
of the gas cloud around the BS (S4, left)
and the Schwarzschild BH (S5, right).}
     \label{fig:snaps0till100CGT11}
\end{figure*}
\begin{figure*}[p!]
  \includegraphics[clip,width=0.8\linewidth]{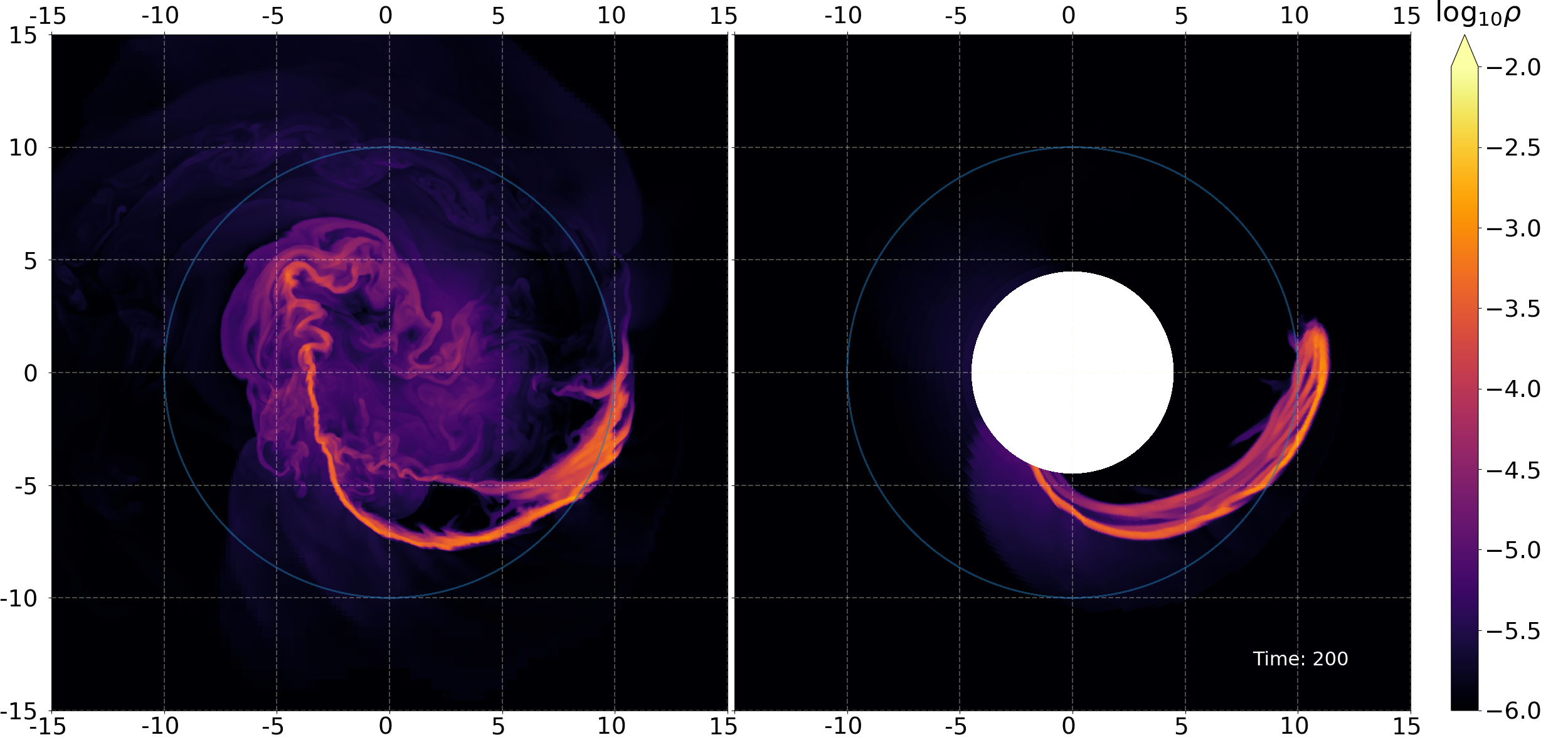}
  \includegraphics[clip,width=0.8\linewidth]{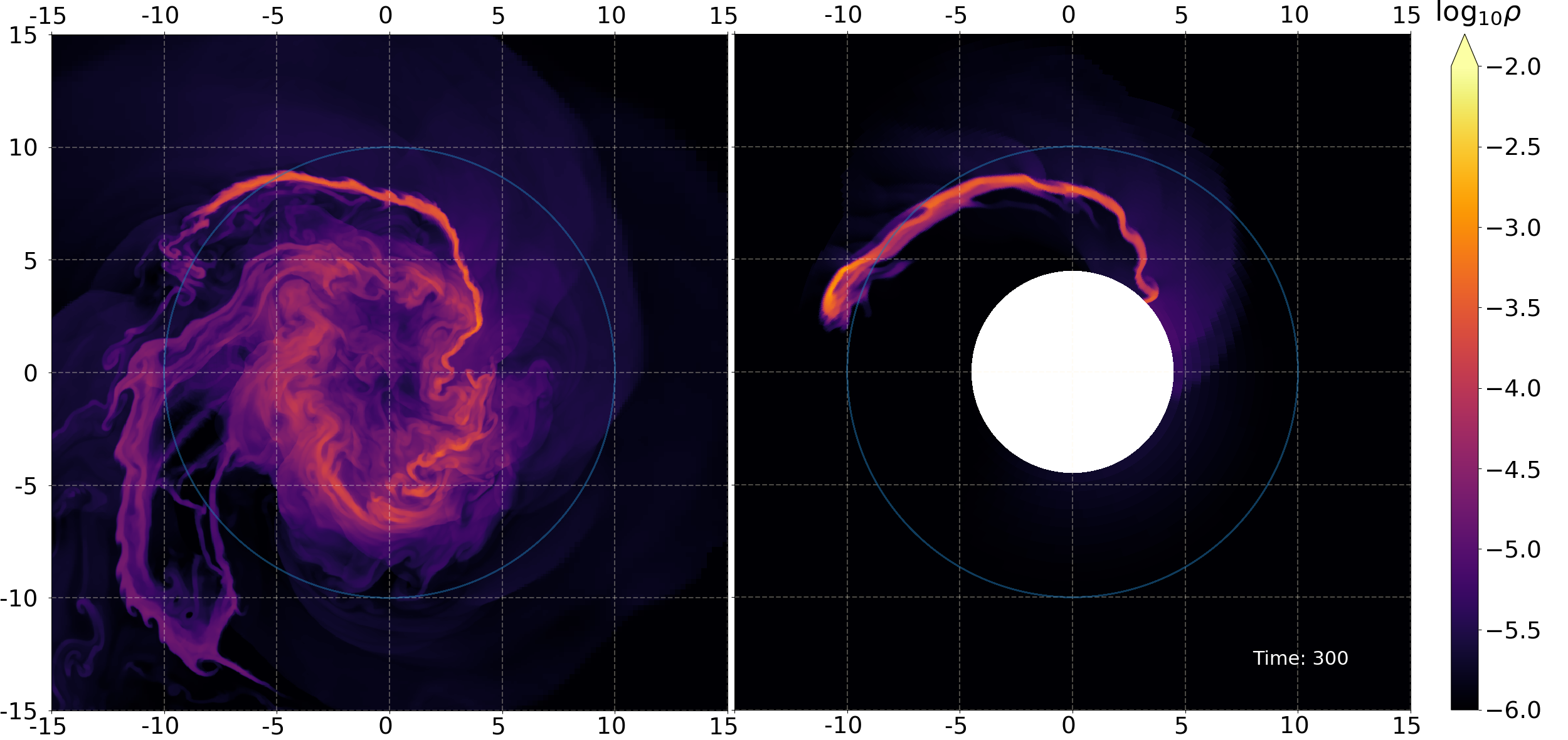}
  \includegraphics[clip,width=0.8\linewidth]{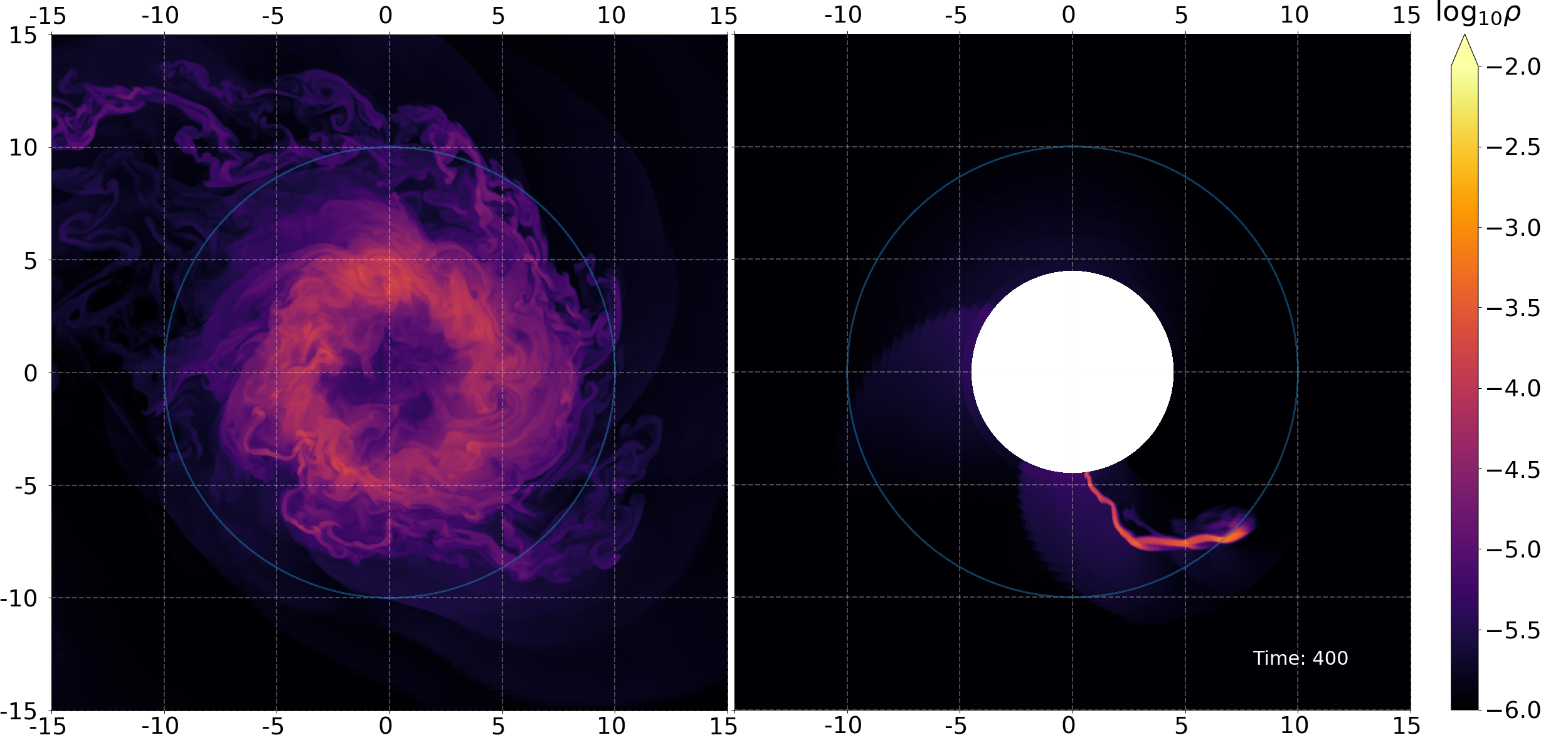}%
     \caption{Simulations S4 and S5:
Selected snapshots at time $t=200$, $300$ and $400$
of the density (with logarithmic color coding)
for an (initially) circular orbit 
of the gas cloud around the BS (S4, left)
and the Schwarzschild BH (S5, right).}
     \label{fig:snaps200till400CGT11}
\end{figure*}

\begin{figure}[h!]
  \centering
{  \includegraphics[width=0.9\columnwidth]{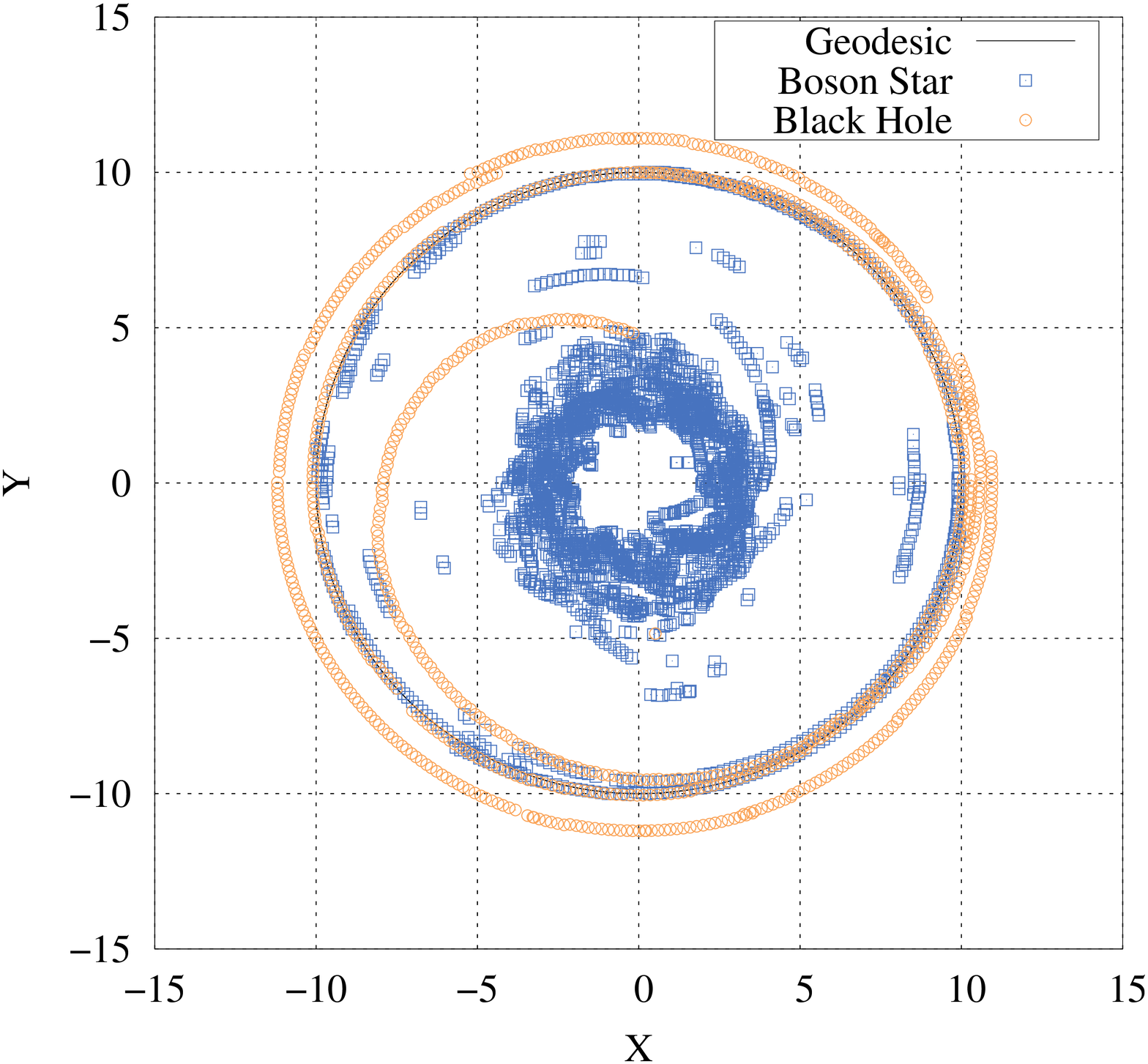}}
{  \includegraphics[width=0.9\columnwidth]{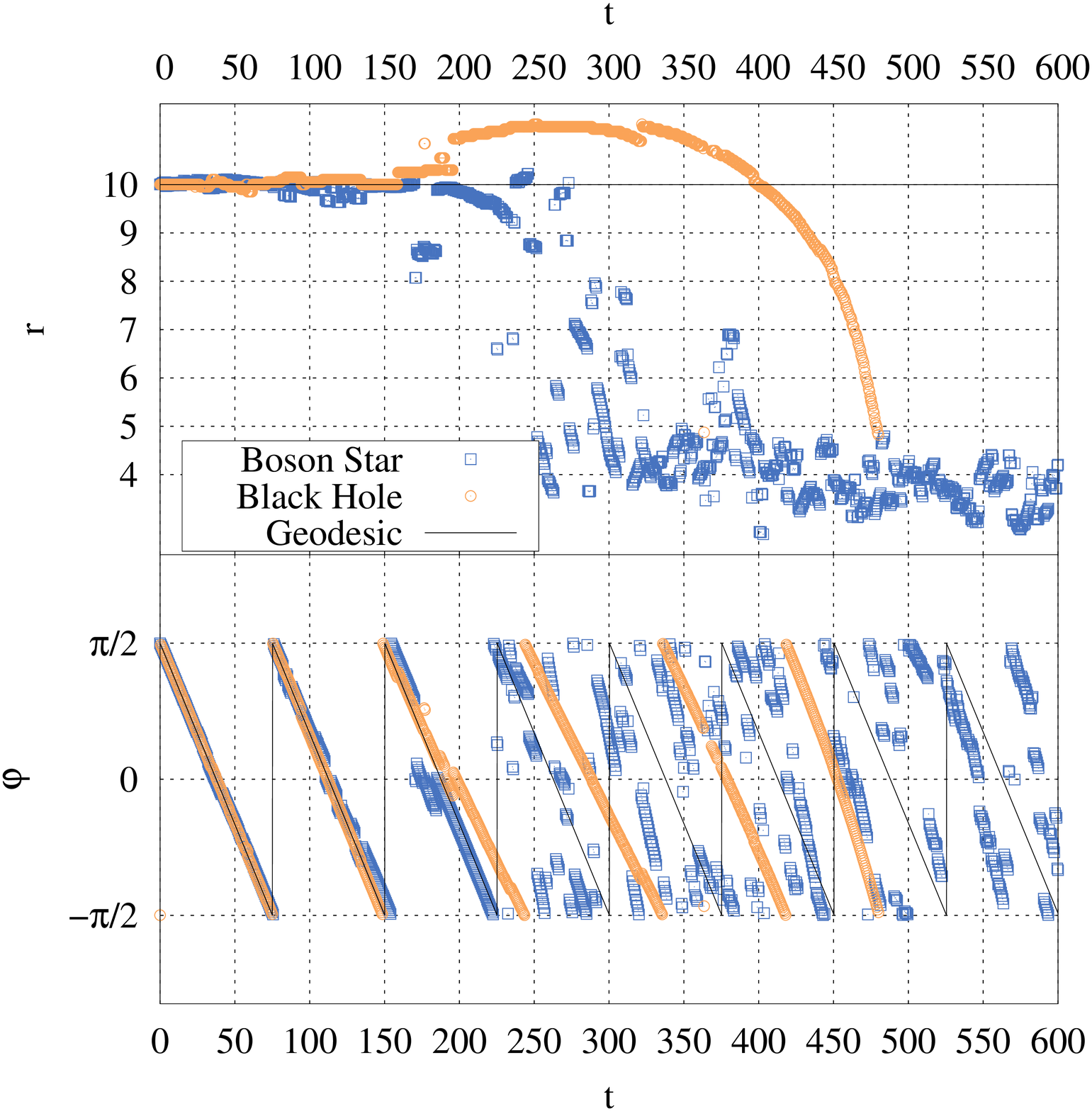}}
     \caption{Simulations S4 and S5:
Position $(x,y)$ (left) and 
$(r,\varphi)$ vs time $t$ (right)
of the maximum density $\rho_{\rm max}$ of the cloud
around the BS (blue) and the Schwarzschild BH (orange)
for an (initially) circular orbit,
together with the corresponding test particle geodesic (black).}
     \label{fig:geoCGT11}
\end{figure}

A set of selected snapshots of the gas density,
in logarithmic scale, is shown in
Fig.~\ref{fig:snaps0till100CGT11} and
Fig.~\ref{fig:snaps200till400CGT11}
for the two simulations.
They display the initial accumulation
of less dense (but high pressure) fluid
at the center of the BS.
The denser gas is then seen to start forming a turbulent 
disk-like structure around the BS,
where centrifugal forces keep the denser gas
from falling towards the BS center.
The disk-like structure is well developed for $t=400$,
as seen in Fig.~\ref{fig:snaps200till400CGT11}.
This structure seems to become stable and less turbulent
towards the end of the simulation.
Spiralling shock waves are formed during this process,
and keep on going until the end of the simulation.
This endphase of the BS simulation S4 contrasts
completely with the endphase of the BH simulation S5,
where the gas more or less disappears
being swallowed by the BH.

Tracking the position of the maximum density 
for both simulations, as shown in Fig.~\ref{fig:geoCGT11},
it is possible to glean insight concerning the general output 
of the simulations, and their comparison with the test particle orbit. 
Before $t=150$, which represents approximately 
the period of the orbit, the maximum density $\rho_{\rm max}$
follows the geodesic in both simulations. 
For the BS the trajectory then starts to deviate
from the circular orbit due to gas-gas interaction
and the competition with new dense spots that emerge on the grid. 
For the BH the gas can be radially stretched freely;
and from $t=150$ to $t=400$ 
the position of the maximum density even exceeds 
the radius of the circular orbit. 
At this stage the gravitational attraction
finally takes over, 
pulling the entire cloud towards the event horizon.
This happens in such a way that after $t=500$ 
almost no gas is left on the grid. 
Indeed such an outcome is expected 
since for this spacetime the cloud starts partially inside the ISCO of the BH, 
which is located at the radius $R_\textrm{ISCO}=10.5084$.

The BS simulation on the other hand continues beyond this time,
since the gas keeps orbiting 
the center of the BS due to the absence of 
an event horizon or a hard surface. 
Therefore we have continued the simulation until $t=1500$.
  
\begin{figure}[t!]
  \centering
\hspace{0.05cm}{  \includegraphics[width=0.9\columnwidth]{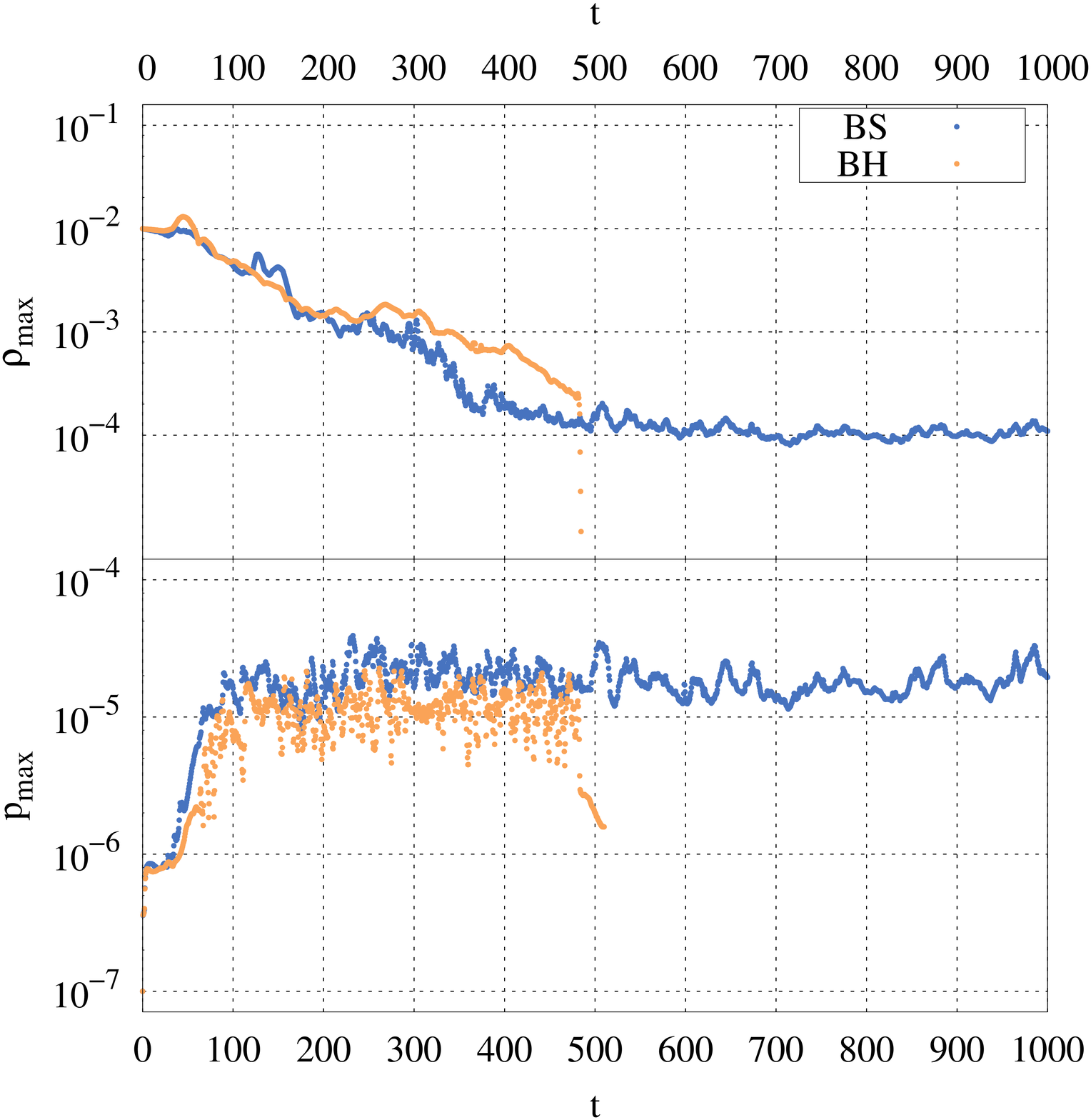}}
{  \includegraphics[width=0.9\columnwidth]{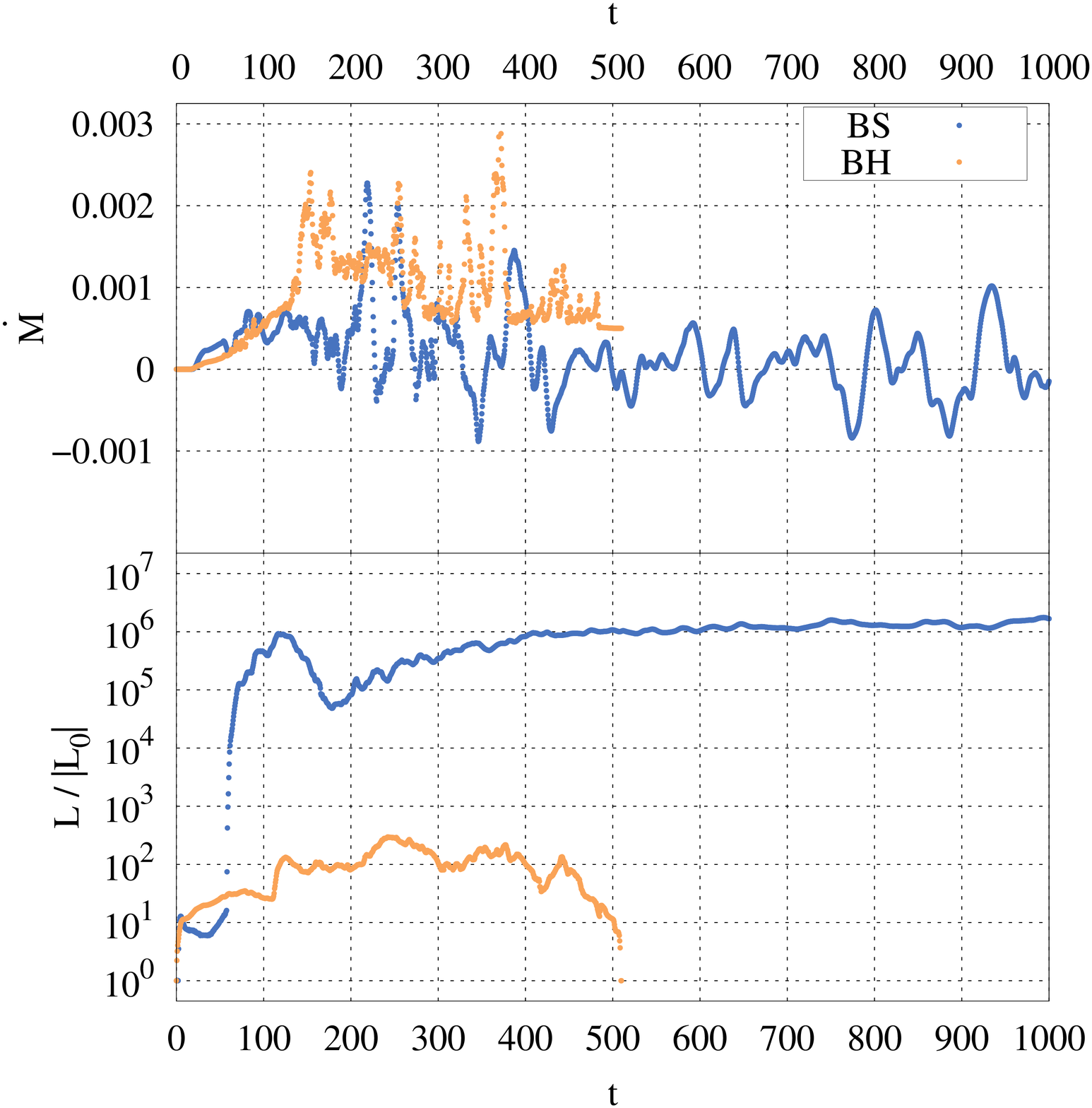}}
     \caption{Simulations S4 and S5:
(Upper panel) Maximum density $\rho_{\rm max}$
and maximum pressure $p_{\rm max}$ vs time $t$
of the cloud
around the BS (blue) and the Schwarzschild BH (orange)
for an (initially) circular orbit.
(Upper panel) Mass flux $\dot M$ and 
normalized total luminosity $L/L_0$
for the BS (blue) and the Schwarzschild BH (orange).}
     \label{fig:prho_accCGT11}
\end{figure}


The global variables of interest are shown in
Fig.~\ref{fig:prho_accCGT11}.
The decrease of the maximum density $\rho_{\rm max}$ 
is similar for both spacetimes for a long time.
Thus after the disruption initiates, which results
from the combination of transverse squeezing
and spaghettification, the fluid does not
get significantly compressed.
The maximum density then stabilizes for the BS simulation,
indicating the existence of a final steady state of the fluid.
Analogously, for both simulations
the maximum pressure $p_{\rm max}$ is seen to first increase
and then oscillate rapidly, assuming similar values in S4 and S5.
However, even though similar in behavior, 
the nature and the consequences of large pressure values
are different for the two spacetimes.

Whereas for the BH spacetime 
high pressure values remain localized,
being only related to the transverse squeezing 
of the cloud and initial non-vanishing velocities, 
for the BS spacetime high pressure values reside
in a more extended region of the fluid.
This region is formed of cloud debris, 
which floats around the center of the grid. 
In fact, this region contributes considerably 
to the total luminosity of the fluid. 
Already during the first stage of the simulation, around $t=70$, 
the first debris of the cloud reaches the center of the BS 
and the high pressure region is formed. 

This region, although initially not very dense,
but featuring high velocities (up to $v^r=0.84$), 
then generates a peak in the total luminosity around $t=120$.
This is seen in Fig.~\ref{fig:prho_accCGT11}(b),
where besides the maximum density and pressure
also the luminosity $L$ (normalized to its initial value $L_0$)
is shown together with the mass flux $\dot{M}$.
After $t=400$ the value of the luminosity from the first peak 
is again restored, staying almost constant now
due to the, now denser, high-pressure region. 
This constant total luminosity is four orders of magnitude bigger 
than the highest luminosity found in the BH simulation.  

Regarding the mass flux $\dot{M}$, as expected, 
only positive values of $\dot{M}$ arise in the BH simulation,
since it represents the accretion rate in this spacetime. 
In contrast, in the BS simulation $\dot{M}$ 
has also negative values.
This does, however, not imply the existence of outflows.

\subsection{Aftermath}

\begin{figure*}[t!]
  \includegraphics[clip,width=0.9\linewidth]{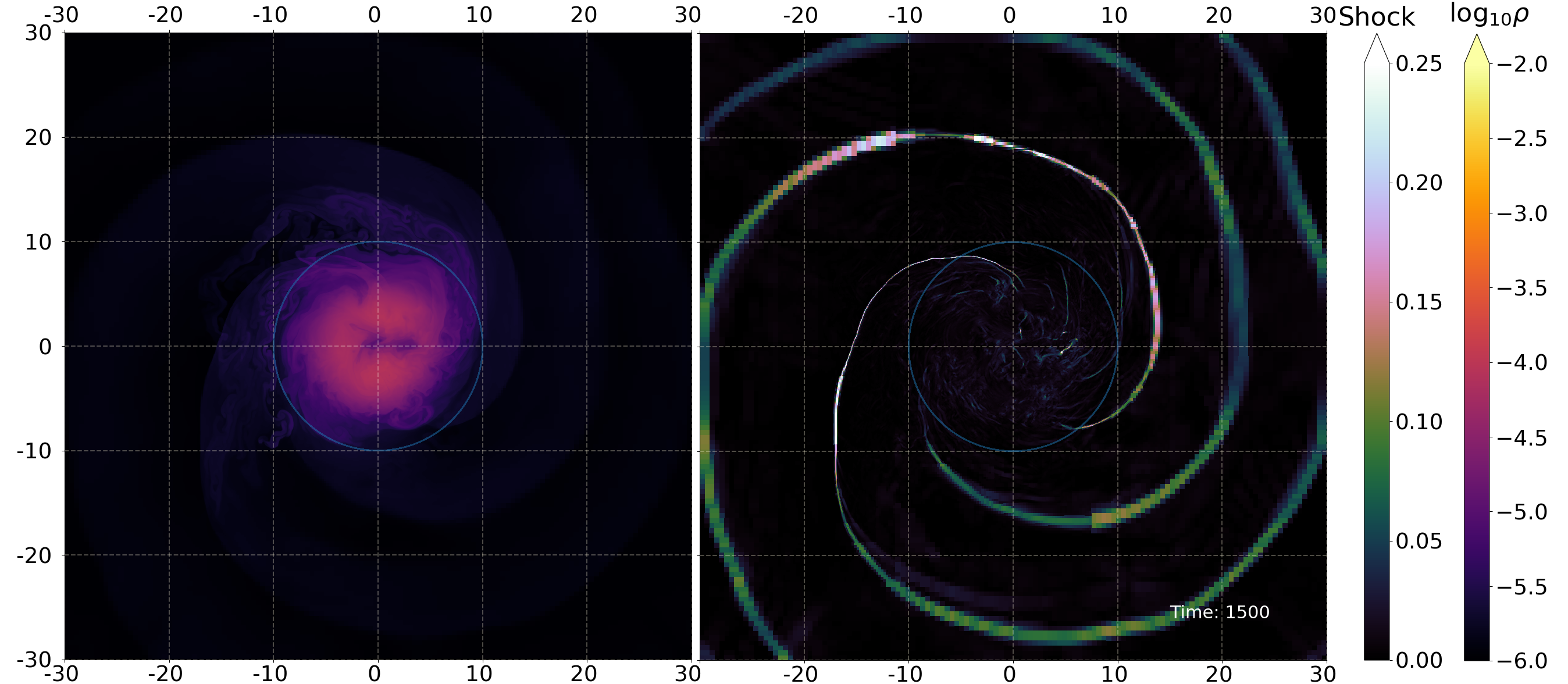}%
     \caption{Simulation S4 aftermath at time $t=1500$:
Snapshot of the density (with logarithmic color coding)
for an (initially) circular orbit 
of the gas cloud around the BS (left), 
and snapshot of the shock detector variable (right).}
     \label{fig:shockandrho1500}
\end{figure*}


\begin{figure}[]
  \centering
  {
  \includegraphics[width=0.9\columnwidth]{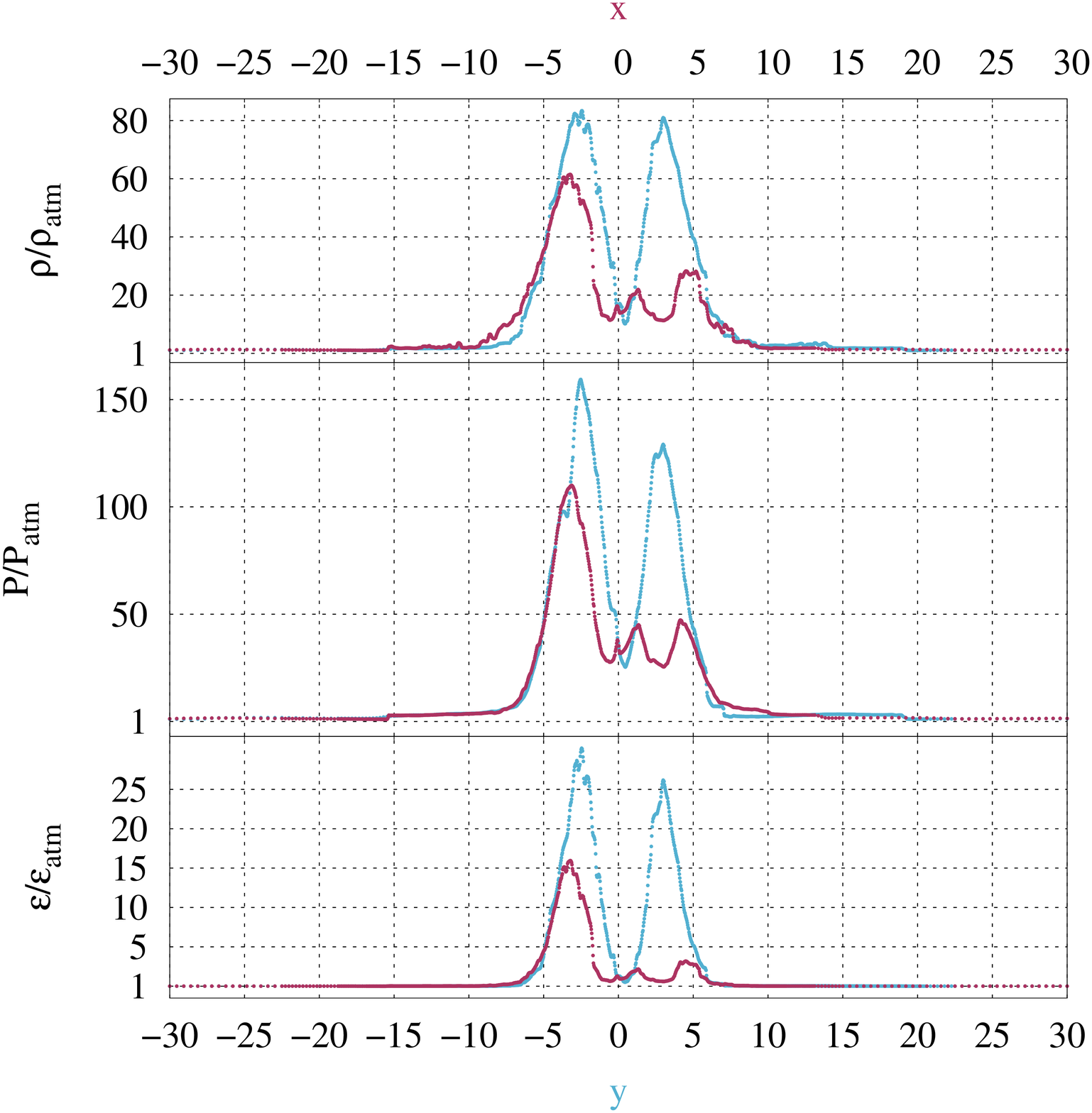}}
 {
  \includegraphics[width=0.9\columnwidth]{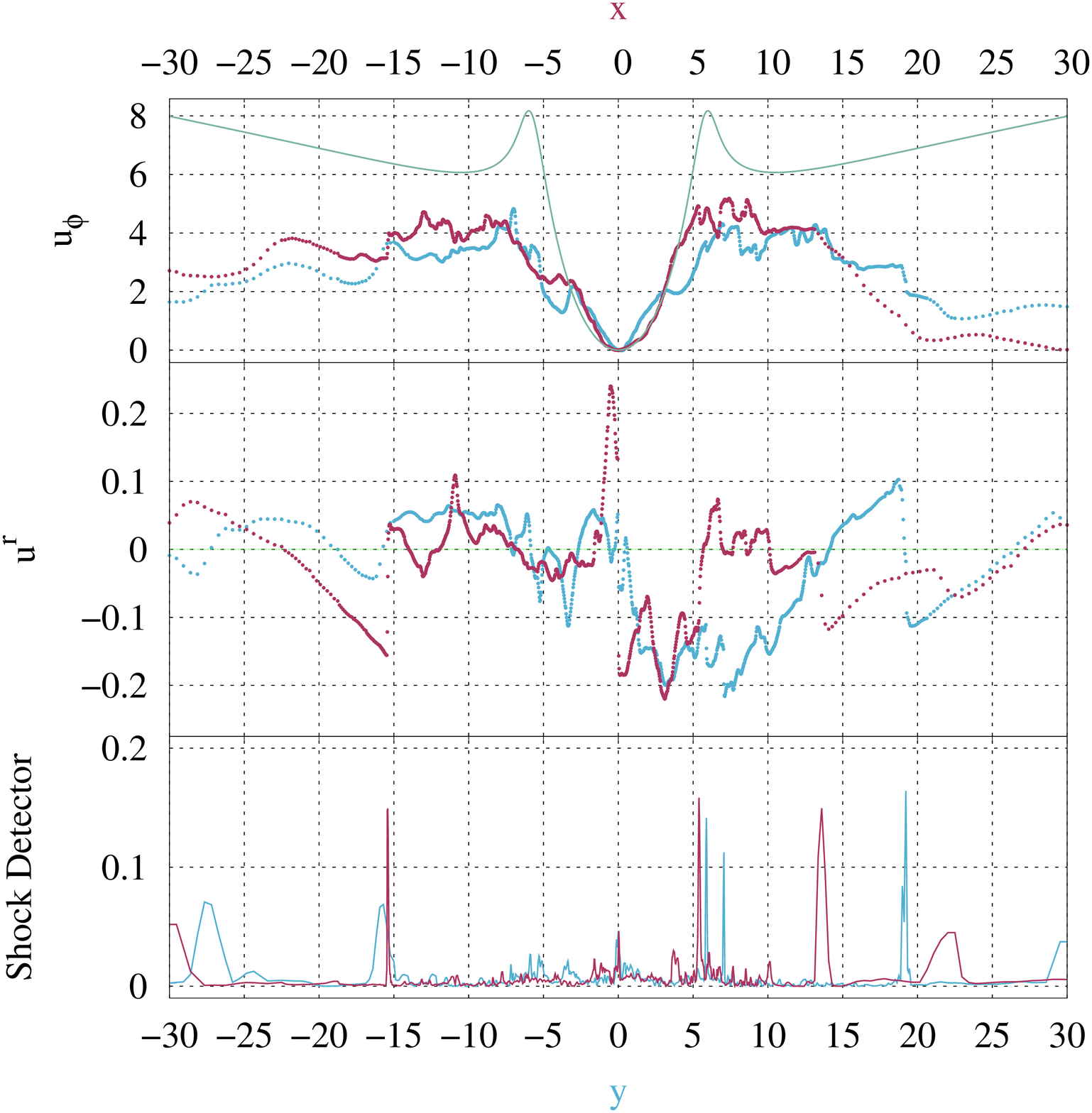}}
 
 \caption{Simulation S4 aftermath at time $t=1500$:
(a - Upper panel) Density profile $\rho/\rho_{\rm atm}$,
pressure profile $p/p_{\rm atm}$,
and emissivity profile $\varepsilon/\varepsilon_{\rm atm}$
along the $x$-axis (red) and the $y$-axis (blue),
normalized with respect to the atmosphere values.
(b - Lower panel) Angular momentum profile $u_\varphi$,
radial velocity profile $u^r$,
and shock detector profile
along the $x$-axis (red) and the $y$-axis (blue).
Also shown are the Keplerian profiles 
for $u_\varphi$ and  $u^r$ (green).}
     \label{fig:slices1500}
\end{figure}

Turning now to the aftermath of the simulations,
we recall that in the BS simulation a disk-like
structure has formed, featuring spiralling shock waves.
An example of these waves together with
the final density configuration of the gas
is shown in Fig.~\ref{fig:shockandrho1500},
representing the final outcome of the simulation S4.

In order to explore the structure and formation 
of the shock pattern let us consider slices of the grid
along the $x$- and $y$-axis at $t=1500$, 
the final time-step of the simulation S4.
These slices are shown in Fig.~\ref{fig:slices1500}(a)
for the density profile $\rho/\rho_{\rm atm}$,
the pressure profile $p/p_{\rm atm}$,
and emissivity profile $\varepsilon/\varepsilon_{\rm atm}$,
normalized by the respective atmospheric values,
and in Fig.~\ref{fig:slices1500}(b)
for the angular momentum profile $u_\varphi$,
the radial velocity profile $u^r$, 
and the shock detector profile,
always for the $x$-axis (red) and the $y$-axis (blue).

\begin{figure}[]
  \centering
  {\includegraphics[width=0.9\columnwidth]{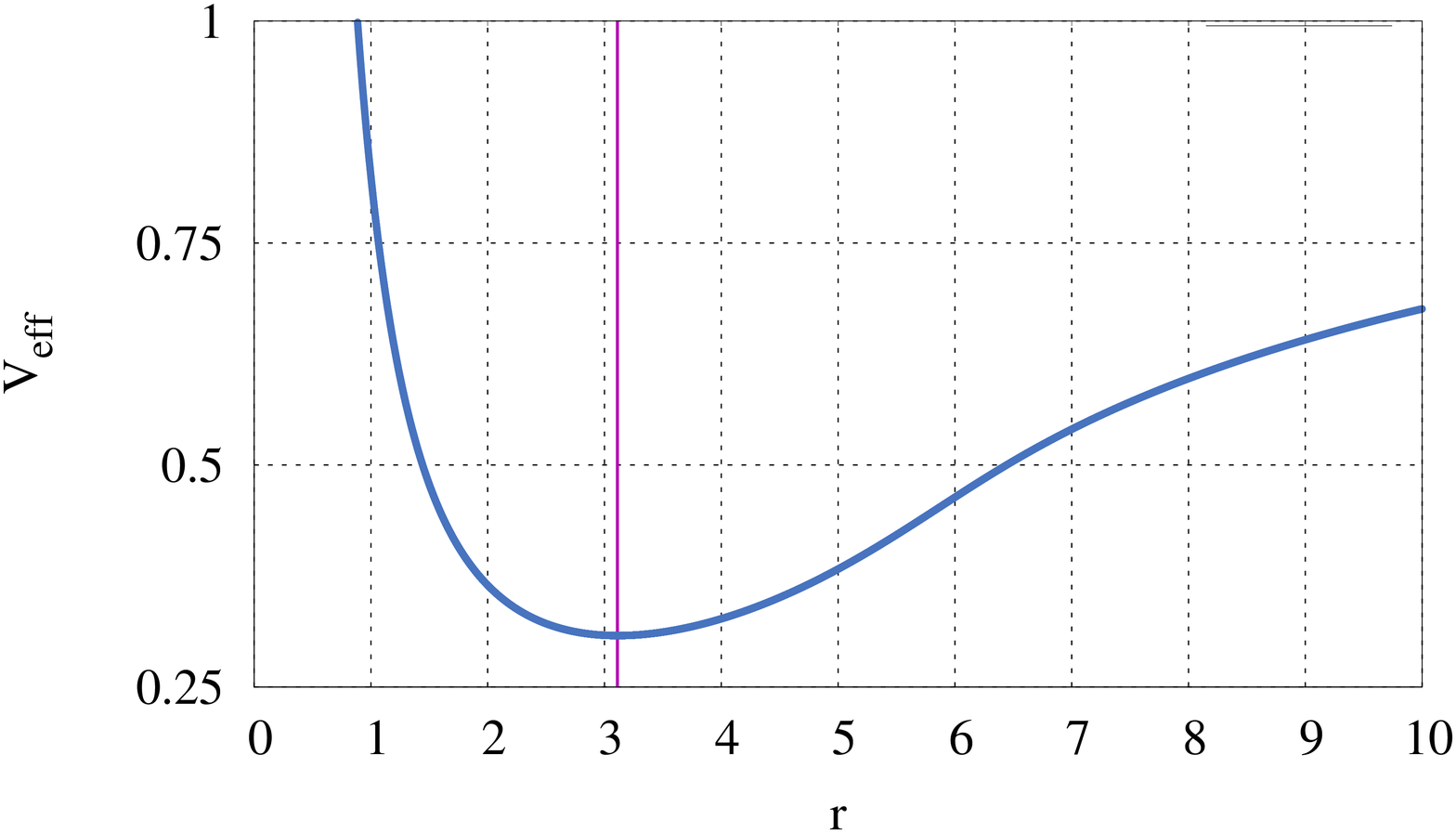}}
  \caption{Effective potential $V_{\rm eff}$ for a massive test particle with angular momentum $L=2$. The purple line at $r=3.1$ marks the position of the circular orbit.}
     \label{potential}
\end{figure}

In Fig.~\ref{fig:slices1500}(a) we observe
almost symmetric configurations for the density,
the pressure and the emissivity along the $y$-axis 
with respect to the BS center.
In contrast, along the $x$-axis the configurations 
do not exhibit such a symmetry.
Whereas the locations of the peaks in $x$
agree with those in $y$ for negative $x$,
we observe much lower values for positive $x$ 
than for negative $x$ for these quantities.
Thus we find a residual low density region
on its way towards the center of the grid. 
While this shows how dynamical the aftermath structure is, 
the balance of centrifugal and gravitational forces 
keeps a ring-like structure roughly in place, 
centered at a radial coordinate $2.9\pm 0.2$
(i.e., $2.9$ is the average of the radial coordinates of the maxima of the density, pressure and emissivity profiles at $t=1500$, and $\pm 0.2$ represents their variation).
By averaging the values of the angular momentum of the cells in the region $2.9-0.2< r < 2.9+0.2$ 
we obtain  $\bar L=\bar{u}_{\varphi}=2$.
Defining the effective potential $V_{\rm eff}$ of a massive test particle via

\begin{equation}
V_{\rm eff}= - g_{tt} \left( 1 + \frac{L^2}{r^2} \right) 
\label{veff}
\end{equation}
we now consider circular geodesics, for which
${V'}_{\rm eff}=0$.
The circular geodesic with angular momentum
$\bar L=\bar{u}_{\varphi}=2$ then resides at 
radial coordinate $r=3.1$,
i.e., in the direct vicinity of the ring-like structure.
Thus the center of the ring-like structure roughly follows a Keplerian geodesic.
The associated effective potential $V_{\rm eff}$ for a test particle with angular momentum $L=2$ is depicted in Fig.~\ref{potential}.

The shock structure spiralling outwards
accompanying the ring 
can be understood better with
the help of Fig.~\ref{fig:slices1500}(b).
Here we see that the angular momentum distribution of the gas 
also appears to be symmetric. 
Note, that besides the data we here also provide 
the Keplerian angular momenta from eq.~(\ref{eq1}) (green).
The gas motion is then, as expected, not geodesic. The gas has super-Keplerian angular momentum 
near the center of the grid,
but becomes sub-Keplerian outside the radius of the ring. 
The $u^r$ velocities, although small, on the other hand 
are not symmetric and feature positive and negative values. 
This reveals how the balance of accreting
and out-flowing gas is working.

Interestingly, by comparing these profiles with the shock detector 
it is possible to infer the origin of the shocks. 
The positions of the shocks outside the ring coincide 
with the positions where the radial velocities change abruptly.
Indeed, the shocks are formed by the colliding gas 
moving outwards (due to the centrifugal forces) 
with the gas moving inwards
(due to the gravitational pull). The shocks then have a radial nature, but are dragged along with the rotating gas creating the spiral structure.
When the dragged gas gets far from the BS center, 
it starts being pulled back towards the center, 
until it reaches another shock wave. 
Although these collisions do not generate radial velocities 
strong enough to destroy the ring, 
they keep generating the shock waves. Due to the nature of the shocks, a meaningful amount of shear is intrinsic to them. Thus this structure is susceptible to cease when viscous time-scales are reached. Nevertheless, due to the highly dynamical scenario this consideration might depend on the nature of the viscosity.
 
We note that, by the end of the simulation S4, 
enough gas from the cloud and its surroundings 
has been spread to the entire grid in a manner 
that no atmosphere is left. 
Therefore the shock waves do not represent an artefact 
of the atmospheric treatment.  We also note that the temperature 
of the gas increases substantially after the disruption, indicating the possibility of ionization. 
It is possible to infer then that the disk formed would become hot plasma, 
for which then magnetic fields would be important. 

In order to address the fate of the disk-like structure, viscous effects would need to be taken into account. Also, depending on the BS model, accumulation of gas at the BS center from accreting matter of the disk might occur \cite{Olivares:2018abq}.

In closing let us mention 
that we have also performed another simulation, 
employing the same parameters as for simulation S4, 
but with vanishing initial angular momentum. 
In this case, although very strong radial shock waves 
are found during the first collision of the gas with the BS, 
after a time of $t=600$ only minor shock waves are formed, 
indicating that the rotation of the gas plays an important role 
in the shock wave dynamics. 
Because of the large extent of the cloud 
and the high velocities involved in such a scenario, 
the fluid oscillation around the BS center 
becomes rapidly turbulent, and the cloud's shape is quickly lost. 
In the aftermath, apart from small vortices 
generated during the first collisions, 
the gas resides symmetrically around the BS 
with a single peak of maximum density and pressure. 
This highly symmetric configuration is reached after $t=600$. 

\section{Conclusions}

In this paper we have reported simulations 
regarding the motion 
and evolution of gas clouds around 
compact spherically symmetric BSs.
We have compared with the motion
of particles on geodesics,
and have investigated the tidal 
and hydrodynamic effects on the
evolution and final disruption
of these clouds. In fact, effects like debris formation and disruption have the tendency to divert the cloud's gas from corresponding particle geodesic motion.
In particular, we have considered 
two different regimes. 
The first regime is the one of dense gas spots nearby the BSs. 

These nearby clouds feature different debris formation mechanisms,
depending on the initial angular momentum provided to them. Namely, the more elliptical the cloud's initial orbit, the stronger is the debris formation and the observed subsequent disruption, for these effects are more significant when the cloud is susceptible to radial motion. Although the cloud's center is inclined to follow the particle geodesic motion, these effects tend to deform the clouds' initial shape and divert portions of the gas from the initial orbit. Particularly for the case in which these small clouds are set out to travel at a constant radius the tidal forces are less significant. On the other hand, the fluid motion through the medium 
gives rise to a prominent turbulent tail. 
This turbulence is related to the fact 
that debris-cloud collisions are still observed. 
We conclude from the performed simulations
that small clouds nearby spherically symmetric BSs 
are less stable and thus possess a shorter lifetime
than the ones in the vicinity of rotating BSs,
studied in \cite{Meliani:2017ktw}.

In a second regime we have provided simulations 
aimed at a comparison between extended clouds 
in a BS spacetime and a Schwarzschild BH spacetime,
with both compact objects possessing the same mass.
Due to their extension, tidal forces rapidly divert portions of the cloud from their initial orbit, enabling the gas to reach the vicinity of the central objects.
Initially in a circular orbit, 
the clouds are seen to behave in a similar manner 
in the initial phase of the simulations. 
But soon important differences arise. 
Most importantly, due to the absence of an event horizon 
in the BS simulation,
the debris of the cloud does not disappear when accreted.
Therefore thermal Bremsstrahlung emissivity is
considerably higher and lasts much longer in a BS spacetime. 
In contrast, in a BH spacetime the gas of the cloud 
in a close-by circular orbit is totally swallowed by the BH. 
The absence of an event horizon or a hard surface 
provides an appropriate environment for the gas 
to stabilize in a ring-like structure
with a spiralling shock wave pattern. 
In contrast, for a BH spacetime no such structure is formed. We believe that a further analysis of these structures regarding their possible similarities with accretion disk models as well as the effects of viscosity on them is an appealing field for future work.


Here we have selected a stable BS with high compactness in order to be rather close to the BH limit in the simulations. However, most BS solutions of the model feature lower compactness, as seen in Fig.~\ref{mxr_bs}. When we choose a less compact BS on the stable branch, but retain comparable conditions for the initial properties of the cloud, we observe qualitatively similar outcomes, as indicated by preliminary simulations. In particular, we retain the formation of a ring-like structure in the aftermath, whose center follows roughly a Keplerian orbit.

In this work we have also provided a preliminary analysis 
regarding tidal disruption events
around compact spherically symmetric BSs. 
Indeed, such spacetimes turn out to be appealing scenarios 
for interesting fluid dynamics-related events to arise. 
Inspired by these first simulations, 
as future work 
we shall aim to perform further simulations 
for different types of BSs, as well as for other
exotic compact objects. 
In particular, regarding rotating spacetimes, 
we think that performing fully 3D simulations, 
as well as including magnetic-fields and accretion disks 
would be fruitful.


\section*{Acknowledgements}

We would like to express our gratitude to
Hector Olivares for introducing us to the numerical code BHAC,
and for his many helpful suggestions with respect to 
the usage of BHAC.
We gratefully acknowledge support by the DFG funded
Research Training Group 1620 ``Models of Gravity''.
LGC would like to acknowledge support via an
Emmy Noether Research Group funded by the DFG
under Grant No. DO 1771/1-1.
The authors would also like to acknowledge networking 
support by the COST Actions CA16104 and CA15117.
The simulations were performed on the HPC Cluster CARL 
funded by the DFG under INST 184/157-1 FUGG.



\end{document}